\documentclass[acmsmall]{acmart}
\usepackage{graphicx}
\usepackage{subcaption}
\usepackage{makecell}
\usepackage{tikz}
\usepackage[table]{xcolor}
\newcommand{\filledcircled}[1]{%
  \tikz[baseline=(char.base)]{
    \node[shape=circle, fill=black, text=white, inner sep=2pt] (char) {#1};}}

\AtBeginDocument{%
  }

\usepackage{ulem} 

\begin{document}

\title{Efficient Page Migration in Hybrid Memory Systems}


\author{Upasna}
\affiliation{%
 \institution{Indian Institute of Technology Ropar}
 \city{Ropar}
 \country{India}}
\email{upasna.21csz0002@iitrpr.ac.in}

\author{Venkata Kalyan Tavva}
\affiliation{%
 \institution{Indian Institute of Technology Ropar}
 \city{Ropar}
 \country{India}}
 \email{kalyantv@iitrpr.ac.in}



\begin{abstract}
Heterogeneous Memory Architecture (HMA) aims to optimize memory usage by leveraging a combination of memory types, such as high-bandwidth memory (HBM), commodity DRAM, and non-volatile memory (NVM), when utilized as main memory. To achieve maximum performance benefits, frequently accessed data pages are prioritized for storage in the faster HBM, while less frequently accessed pages are stored in slower memory types like DRAM or NVM. This enables a more efficient allocation of memory resources and improves overall system performance. In a Flat Address Space memory organization, all memory types, both fast and slow, are treated as a unified memory pool. This approach increases the overall memory capacity accessible to the system. In Flat Address Space organization, frequently accessed data pages may need to be remapped from slower memory to faster memory to improve memory access times. Such relocation requires changes to the data/states in the TLB (TLB shootdown) and the processor's cache (cache line invalidations), leading to performance degradation. To address these inefficiencies, we propose a novel solution called {\tt Duon}. The goal of {\tt Duon} is to eliminate the overheads associated with page migration in systems using Extended TLB and Page Table. Specifically, our approach ensures that the updated mapping information for remapped pages is carefully stored directly in the TLB and page table itself. By doing so, the need for TLB shootdown and cache line invalidation after page migration is eliminated. Consequently, our proposal results in an overall improvement in IPC by 3.87\% over existing state-of-the-art techniques, enhancing the efficiency and performance of heterogeneous memory systems. Further, our approach can work with any of the existing page migration policies and improve the performance.

\end{abstract}

\ccsdesc[500]{Computer systems organization~Heterogeneous (hybrid) systems}


\keywords{Heterogeneous Memory Architecture (HMA), Flat Address Space, Page Migration}


\maketitle

\section{Introduction}
As the demand for memory continues to grow exponentially due to the increasing dependency on data-intensive applications, conventional Dynamic Random Access Memory (DRAM) technology is posing new design challenges due to its limitations. These limitations include increased power consumption, slow latency improvements and rising costs, that have made it difficult to scale DRAM efficiently. This is particularly concerning in a world where cutting-edge applications like graph analytics \cite{GraphAnalytics}, genome sequencing \cite{GenomeSequence}, big data processing \cite{BigData}, Large Language Models (LLMs) \cite{ChatGpt}, Internet of Things (IoT) \cite{IOT}, and cloud computing \cite{CloudComputing} present large memory footprints while simultaneously demanding low power consumption and high data throughput.

In response to these challenges, new alternatives to traditional DRAM have been developed. Among these are advanced 3D-stacked DRAM technologies such as the Hybrid Memory Cube (HMC) \cite{HMC}, HBM \cite{HBM}, and Wide-IO (WIO) \cite{WideIO}. These innovations aim to address some of the bottlenecks of conventional DRAM by offering higher performance and reduced energy consumption. Additionally, Low Power DRAM (LPDRAM) \cite{LPDDR5} has been introduced to cater to devices and systems requiring energy efficiency. However, while these new DRAM technologies provide notable improvements, each design option possess a unique set of limitations, such as limited memory capacities and relatively higher costs, making them less scalable for enterprise/server-scale system demands.

Meanwhile, emerging technologies such as phase change memory (PCM) \cite{PCM}, spin transfer torque RAM (STT-RAM) \cite{STT-RAM}, and Intel's 3D XPoint \cite{3DXPoint} technology, offer promising alternatives to overcome DRAM limitations. NVMs provide low static power, lower cost, high density, and greater capacity, but face challenges like limited write endurance, higher dynamic write energy, read-write asymmetry, and increased latency. To address these trade-offs, modern systems adopt HMA, combining DRAM/HBM performance with NVM/DDR capacity for scalable, balanced memory. By leveraging the strengths of each technology, HMA meets growing memory demands with improved efficiency, scalability, and performance.


\textbf{Our observation:} In heterogeneous memory systems, the coexistence of high-speed, high-cost memory and high-capacity, cost-effective memory provides opportunities to achieve both performance and scalability. This hierarchy may involve combinations such as HBM with NVM (For e.g., PCM), or HBM with traditional DRAM (For e.g., DDR), among others. Effectively managing such diverse memory types requires careful design of the memory sub-system that  can perform data placement and movement through optimal page migration strategies. Page migration refers to the movement of pages (typically of $4KB$ or $8KB$ size) between the fast and slow memories within the HMA. Page migrations are primarily done so as to improve the performance of the HMA by maximizing hits to the faster memory. When a page migration is performed, the upper (closer to the processor) levels of the memory need to be updated. A critical challenge arises when frequent invalidation of TLB entries and cache lines occurs during page migration. This incurs significant system overhead, undermining the potential performance benefits of migration. Additionally, the variation in access latency, bandwidth, and write endurance across memory tiers complicates migration decisions. These challenges highlight the need for a robust mechanism to streamline page migration and reduce its associated overhead. 

\textbf{Our approach:} To address these challenges, we propose a novel solution, namely, {\it {\tt Duon}}. {\tt Duon} extends the traditional TLB and page table structure to include newly remapped physical addresses, at the same time preserving the initial virtual-to-physical address mapping throughout the migration process. By eliminating the need for frequent invalidation operations in the TLB and cache due to small size of separate structure like remap table as proposed in prior techniques \cite{OnTheFly}, {\tt Duon} significantly reduces overhead due to the page migrations. The methodology involves setting an Ongoing Migration Flag to indicate when migration is in progress and employing additional metadata flags to optimize the migration process. These metadata flags, including the Buffer Residency Flag and the Pair Flag, enable efficient management of data movement between different memory types, ensuring that frequently accessed data resides in faster memory while leveraging the capacity of slower memory for less critical data. It can be noted that {\tt Duon} does not propose a new migration strategy, rather addresses the fundamental side effect with page migration.\\

\textbf{Our contribution:} The key contributions of this work are as follows:
\begin{itemize}
    \item \textbf{Novel Page Table Extension:} We introduce {\tt Duon}\footnote{{\tt Duon} means {\it pair}, it broadly covers the idea of keeping two mappings i.e., initial physical address and remapped physical address corresponding to a virtual address.}, a mechanism that augments the traditional TLB and page table to accommodate remapped physical addresses, reducing the need for frequent invalidation operations and minimizing performance degradation during page migration (Section~ \ref{our_idea}). {\tt Duon} can work with any of the existing page migration strategies.

    \item \textbf{Performance Optimization in Heterogeneous Systems:} {\tt Duon} allows data placement decisions to benefit from the characteristics of both faster and slower memory, by minimizing disruption from page migration overheads. {\tt Duon} demonstrates a significant performance enhancement in terms of Instructions Per Cycle (IPC) compared to existing techniques. Our experimental evaluation (Section~ \ref{results_analysis}) shows that {\tt Duon} achieves performance improvement as high as 13.39\%, showcasing the system's ability to optimize performance in favorable scenarios. Overall, across a range of workloads, {\tt Duon} delivers an average performance improvement of 3.87\%, reflecting consistent and measurable gains in efficiency. We demonstrate that {\tt Duon} provides a scalable solution to manage the increasing complexity of heterogeneous memory architectures (Section~ \ref{sensitivity_analysis}), paving the way for improved system performance and cost efficiency in future computing systems.

\end{itemize}

\textbf{Organization of the Paper:} The rest of the paper is organized as follows: Section~\ref{background} provides the background. Section~\ref{PriorWork} discusses prior work. Section~\ref{motive} describes the motivation for the proposed work, delving into the challenges of page migration overhead, establishing the need for a novel approach. Section~\ref{our_idea} contains the proposed innovative approach, namely {\tt Duon}, detailing its design, working, and theoretical aspects. Section~\ref{exp_setup} presents the experimental setup that outlines the methodology, tools, configurations, and benchmarks used to evaluate the proposed solution. Section~\ref{results_analysis} presents the experimental findings, which demonstrate the effectiveness of {\tt Duon} and provides a discussion on its impact on performance compared to the state-of-the-art approaches. Finally, Section~\ref{conclude} concludes and summarizes the key insights and implications of the research.

\section{Background}
\label{background}

\begin{figure}[h]
  \centering
  \includegraphics[width=0.6\linewidth, keepaspectratio, height=4cm]{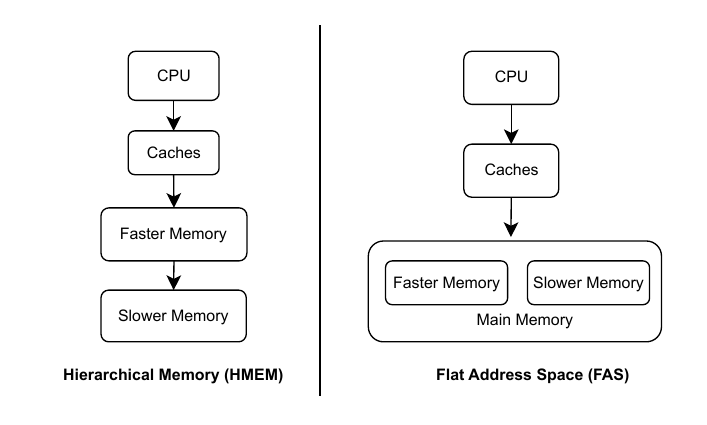}
  \caption{Overview of two design choices of Heterogeneous Memory.}
  \label{Hierarch_Flat_HMA}
  \Description{Block diagram showing two design choices of Heterogeneous Memory}
\end{figure}

HMA design choices can be broadly classified into two types: Hierarchical Memory design (HMEM) and Flat Address Space design (FAS)\footnote{We use the abbreviations HMEM and FAS for Hierarchical Memory Design and Flat Address Space Design respectively throughout the paper.} as shown in Fig.~\ref{Hierarch_Flat_HMA}. Each of these designs has its own strengths, weaknesses, and application scenarios, making them suitable for different types of workloads. In HMEM, faster memory (e.g., DRAM or HBM) acts as a cache for slower memory (e.g., NVM or DDR4/DDR5). This setup boosts performance by storing frequently accessed data in fast memory, managed at hardware level without major OS changes. However, it requires metadata like tag stores, introducing overhead in memory usage and latency. Nonetheless, hardware management in this approach alleviates the need
for software-level intervention, making it relatively seamless to implement in existing systems.

On the other hand, FAS takes a different approach by allowing both the faster and slower memories to be accessed concurrently. FAS treats all memory as part of a unified address space. This design offers distinct advantages for workloads that require large memory capacities, as it avoids duplicating data in the faster memory. The choice between the two designs often depends on the characteristics of the workloads. HMEM is highly advantageous for workloads that exhibit high spatial locality and have memory footprint that fits within the size of the faster memory. Meanwhile, FAS is particularly beneficial for workloads that are constrained by memory capacity.
Recent works propose Hybrid Cache-Flat Memory Design \cite{Hybrid2}, wherein a portion of fast memory operates as a cache, while the remaining is mapped directly into the system's flat address space. This dual-function architecture seeks to combine the performance benefits of caching with the capacity and flexibility of flat memory.

While both HMEM and FAS offer compelling benefits, FAS introduces unique system-level challenges stemming from its reliance on software-driven data movement. In particular, as the OS assumes responsibility for managing page migrations between the faster and slower memories, the cost of maintaining coherence across the system becomes increasingly significant. One critical drawback is the overhead introduced by TLB shootdowns and cache line invalidations, which occur each time a page is migrated. These operations, though essential for correctness, can negate the performance advantages of FAS, especially in multi-core and memory-intensive environments. The following section explores these limitations in greater depth and motivates the need for hardware-assisted solutions to overcome them efficiently.

As the number of cores increases and memory systems grow more complex, the bottlenecks due to page table coherence and TLB maintenance become a key limiter to scalability in heterogeneous systems.  These limitations motivate a redesign of the page table architecture itself, especially in systems where frequent page remapping is necessary. To mitigate migration overheads without sacrificing consistency or correctness, we propose hardware-assisted TLB and page management solution, {\tt Duon}, that can decouple data movement from virtual address semantics. 
{\tt Duon} reimagines page translation through the lens of remap-aware page tables and a hardware-based TLB coherence mechanism. By embedding migration intelligence into the page table structure and leveraging in-memory metadata flags, {\tt Duon} enables efficient, low-overhead, and scalable page movement in flat-address heterogeneous memory systems.

\section{Prior Works}
\label{PriorWork}
Prior work in the domain of heterogeneous memory system can be broadly categorized into three architectural models, each differing in how fast and slow memory tiers are integrated and managed. These models include HMEM, FAS, and Hybrid Cache-Flat Memory Design as discussed in Section~ \ref{background}. 

\subsection{Hierarchical Memory Organization}
To address the trade-offs of HMEM, particularly those involving block granularity, cache associativity, and replacement policies, several designs have been proposed. Examples include LH Cache \cite{LHCache}, Alloy Cache \cite{AlloyCache}, Footprint Cache \cite{FootprintCache}, Unison Cache \cite{UnisonCache}, and ACCORD \cite{Accord}. Each of these designs represents a unique strategy aimed at enhancing the performance and efficiency of DRAM Cache systems under different workload conditions as shown in Table \ref{hard-dram-cache-comp}.

    

\begin{table}[h]
    \centering
    \scriptsize
    \caption{Hardware based DRAM Cache in terms of associativity and block granularity.}
    \label{hard-dram-cache-comp}
    \begin{tabular}{|>{\raggedright\arraybackslash}p{1.5cm}|
                        >{\raggedright\arraybackslash}p{2cm}|
                        >{\raggedright\arraybackslash}p{2cm}|
                        >{\raggedright\arraybackslash}p{2cm}|
                        >{\raggedright\arraybackslash}p{2cm}|
                        >{\raggedright\arraybackslash}p{1.5cm}|}
        \hline
        \rowcolor{gray!=20}
        \textbf{Work} & \textbf{Associativity} & \textbf{Granularity} & \textbf{Advantage} & \textbf{Hit Latency} & \textbf{Miss Latency} \\ \hline
        LH Cache \cite{LHCache}      & Set Associative      & 64 Byte     & Compound Tag and Data Access     & MissMap + DRAM Tag
        & MissMap Lookup     \\ \hline
        Alloy Cache \cite{AlloyCache}      & Direct Mapped      & 64 Byte & TAD Unit, Optimized Hit Latency     & Predictor + DRAM TAD Read
        & Predictor Lookup  \\ \hline
        Footprint Cache \cite{FootprintCache}       & Set Associative      & Allocates at Page granularity, fetches at Cache line granularity      & Reduces Overfetching, Optimizes Bandwidth Consumption
        & SRAM Tag + DRAM Data Read      & SRAM Tag Lookup      \\ \hline
        Unison Cache \cite{UnisonCache}      & Set Associative      & Page Size (2KB)      & Scalable      & Overlapped DRAM Tag + Data Reads
        & DRAM Tag Lookup      \\ \hline

    \end{tabular}

\end{table}

In addition to hardware-centric approaches, software-managed DRAM Cache techniques have been proposed to reduce the storage overhead associated with maintaining the tag-store in DRAM Cache. These techniques aim to embed the cache tags into unused bits of the Page Table Entry (PTE) and cache these tags within the TLB. Notable designs in this category include Tagless \cite{Tagless}, HSCC \cite{HSCC}, and Banshee \cite{Banshee}. However, these OS-managed DRAM Cache designs come with a significant drawback: the application thread experiences a stall whenever there is a tag-miss, which can impact overall system performance. To address this limitation, non-blocking software-managed DRAM Cache designs, such as NOMAD \cite{Nomad} has been developed, which aim to minimize the performance overhead caused by tag-misses. This innovative approach eliminates the need for stalling application threads, thereby improving the efficiency of software-managed caching systems.

\subsection{Hybrid Cache-Flat Memory Design}
This hybrid approach seeks to balance the trade-offs between performance and memory availability. Several innovative techniques have been proposed to maximize the benefits of DRAM Cache while minimizing the loss of capacity. Notable examples include CAMEO \cite{Cameo}, Chameleon \cite{Chameleon}, and Hybrid$^2$ \cite{Hybrid2}, that are broadly compared in Table \ref{soft-dram-cache-comp}. These designs aim to achieve a balance by optimizing the performance of DRAM Cache for speed-critical workloads without significantly compromising the available memory for capacity-bound workloads. By adopting such strategies, these techniques ensure that the memory system can handle a wide range of application demands more effectively.

\begin{table}[h]
    \centering
    \scriptsize
    \caption{Software Managed DRAM Cache.}
    \label{soft-dram-cache-comp}
    \begin{tabular}{|>{\raggedright\arraybackslash}p{2.5cm}|>{\raggedright\arraybackslash}p{4.5cm}|>{\raggedright\arraybackslash}p{4.5cm}|}
        \hline
        \rowcolor{gray!=20}
        \textbf{Work} & \textbf{Focus} & \textbf{Limitation} \\ \hline
        ACCORD \cite{Accord} & High hit rate, low hit latency using probabilistic way steering & Reduced scope of improvement for large cache size \\ \hline
        Chameleon \cite{Chameleon} & Switching memory regions between PoM and Cache mode dynamically & Segment sized blocks may not be beneficial for limited locality workloads \\ \hline
        PageSeer \cite{PageSeer} & Perform segment swapping with substantial lead time by page walk hints & Poor prefetch accuracy for pages with diff. access patterns at diff. times \\ \hline
        Hybrid$^2$ \cite{Hybrid2} & Avoid copying data between Cache and flat address using indirection & Gives away 0.3--5.1\% performance as compared to DRAM Cache designs \\ \hline
        NOMAD \cite{Nomad} & Avoid blocking of application thread on tag miss & PCSHR lookup on both tag-hit and tag-miss \\ \hline
    \end{tabular}

\end{table}

In addition to optimizing memory capacity and performance, research efforts have also concentrated on improving bandwidth utilization and prefetching in Heterogeneous Memory Architectures (HMA). Bandwidth is a critical resource in memory systems, and its efficient use directly impacts overall system performance. For example, BATMAN \cite{Batman} is a notable technique aimed at optimizing bandwidth in HMAs, while PageSeer \cite{PageSeer} focuses on enhancing prefetching mechanisms to improve data access patterns and reduce latency. These advancements enable modern systems to achieve higher levels of efficiency and responsiveness, even under demanding workloads.

A significant challenge faced by HMAs is the limited write endurance of NVM technologies. When NVM is used as a component of HMA, excessive write operations can reduce the lifespan of the NVM and increase write energy consumption. Consequently, a major goal of research in NVM-based HMAs is to minimize the number of writes to NVM, thereby improving its durability and energy efficiency. Techniques such as DFPC (Dynamic Frequent Pattern Compression) \cite{DFPC} and DATACON (DATA CONtent aware PCM writes) \cite{DATACON} have been developed to address these issues. DFPC employs dynamic compression algorithms to reduce the amount of data written to NVM, while DATACON optimizes the content of writes to PCM to further extend its lifespan. These solutions demonstrate significant potential in ensuring the long-term viability and energy efficiency of NVM when integrated into heterogeneous memory systems. \\

\vspace{-0.5cm}
\subsection{Flat Address Space (FAS) Memory Architecture}
\label{on-the-fly-adapt}
Meswani et al. propose an epoch-based page migration technique \cite{HMA-HS}, wherein memory access patterns are monitored over discrete time intervals, or epochs, to identify frequently accessed (hot) pages. During each epoch, hardware counters track page-level access statistics, which are then analyzed at the end of the interval to determine which pages should be migrated to faster die-stacked DRAM. This selective migration improves performance by keeping hot data close to the processor, while less frequently accessed (cold) pages remain in off-package DRAM. The technique balances migration overhead with performance gains by avoiding continuous page shuffling and adapting to temporal locality. Their approach exemplifies a hardware/software co-design, where hardware facilitates efficient profiling and software orchestrates page placement, enabling scalable and adaptive memory management in heterogeneous memory systems.

Recent work in the area of FAS Memory organization by Shashank et al. propose On-the-fly Page Migration and Address Reconciliation \cite{OnTheFly}, Dynamically Adapting Page Migration Policies based on Applications' Memory Access Behaviors \cite{DynAdapThold}, and Subpage Migration in Heterogeneous Memory Systems \cite{SubpageMig}. On-the-fly Page Migration and Address Reconciliation seeks to address the overhead associated with invalidating TLB entries and cache lines during page migration. The key idea behind this approach is to perform page migration operations in the background, making them transparent to the user program. This is achieved using specialized Migration Controller (MigC) hardware, which efficiently manages the movement of data without interrupting the execution of applications. MigC is set up as a pseudo-processor that sends write invalidate requests to ensure all caches write back dirty data and clear their cache lines for specific pages. It can communicate with other caches to send requests and receive acknowledgments, but other caches don't send requests to MigC or wait for its responses. This prevents MigC from contributing to traffic during normal system operations. By eliminating the need for application-level awareness of migration, this technique reduces latency and enhances overall system performance.

Another noteworthy contribution is the introduction of Dynamically Adapting Page Migration Policies. This approach emphasizes the importance of tailoring migration strategies to the specific memory access behaviors of applications. It dynamically determines whether an application is migration-friendly or migration-unfriendly and adjusts the migration policy accordingly. For instance, in cases where migration is beneficial, the technique may increase the frequency of migrations to optimize performance. Conversely, when migration introduces overhead or inefficiencies, it can reduce or even halt migrations altogether. This adaptive mechanism ensures that the migration process remains efficient and context-aware, thereby improving system responsiveness.

A third innovative technique is Subpage Migration in Heterogeneous Memory Systems, which focuses on a more granular approach to data movement. Instead of migrating entire memory pages from slower memory to faster memory, this method targets only the portions of a page that are frequently accessed. By transferring high-access portions (or hot data) to the faster memory while leaving less-accessed portions in the slower memory, Subpage Migration reduces unnecessary data movement and saves memory bandwidth. This targeted migration approach is particularly beneficial for workloads with irregular memory access patterns, as it maximizes the utilization of faster memory resources while minimizing the associated overhead. \\

\vspace{-0.8cm}
\section{Motivation}
\label{motive}

Operating system-based memory management solutions, while flexible, are limited by their software-centric approach. They often involve additional latency when handling frequent page migrations, especially in FAS, where all memory types are treated as a unified pool. The challenges of updating mappings in the TLB and invalidating cache lines during page migration exacerbate these performance penalties. Consequently, such overheads can offset the performance gains that heterogeneous memory systems aim to deliver.

This motivates the need for a hardware-centric solution that eliminates the inefficiencies associated with OS-based page migration. Hardware-based approaches can provide direct and low-latency mechanisms for managing memory mappings, thereby significantly reducing the overhead associated with memory relocation. Furthermore, incorporating support for seamless page migration at the hardware level opens up new possibilities for achieving near-optimal memory access times without compromising system performance.

\begin{figure}[htp]
    \centering
    \begin{subfigure}{0.48\textwidth}
        \centering
        \includegraphics[width=0.8\linewidth]{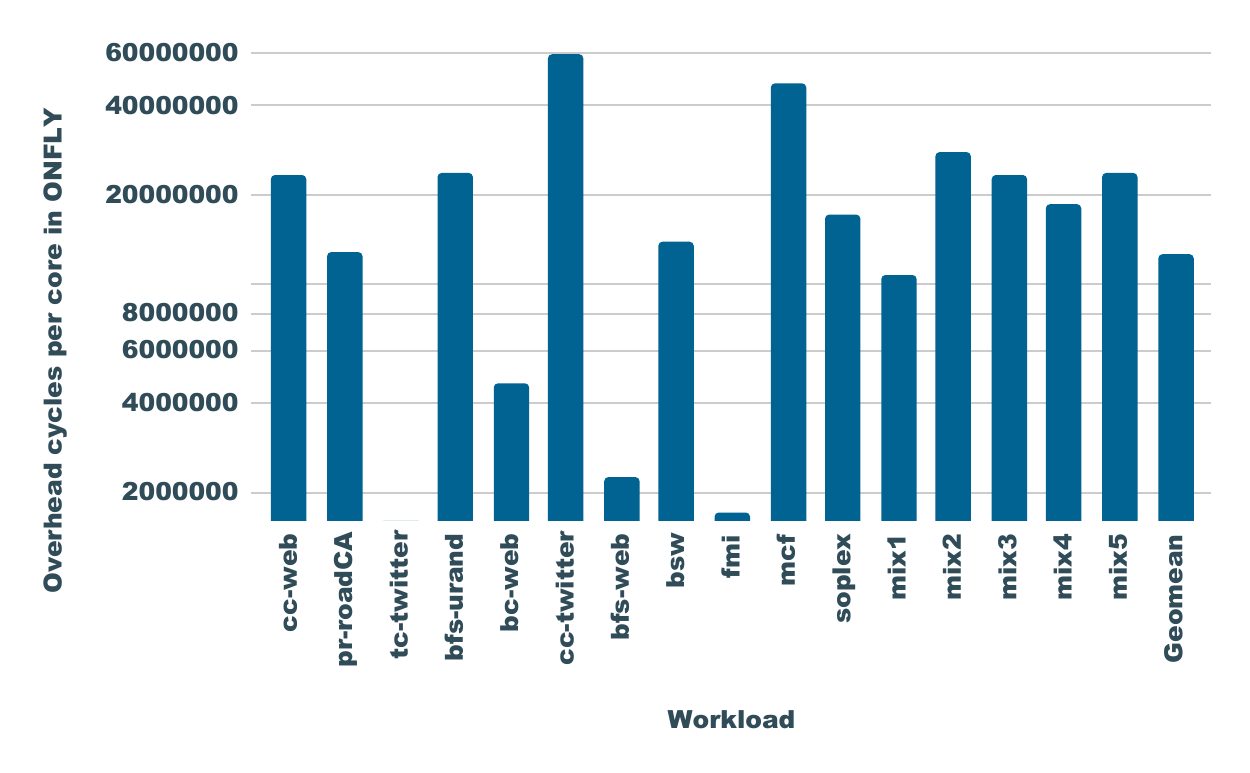}
         \caption{Accumulated overhead cycles per core due to address reconciliation in ONFLY.} 
        \label{hma_onfly_oh_cycles_due_to_ar}
    \end{subfigure}
    \hfill 
    \begin{subfigure}{0.48\textwidth}
        \centering
        \includegraphics[width=0.8\linewidth]{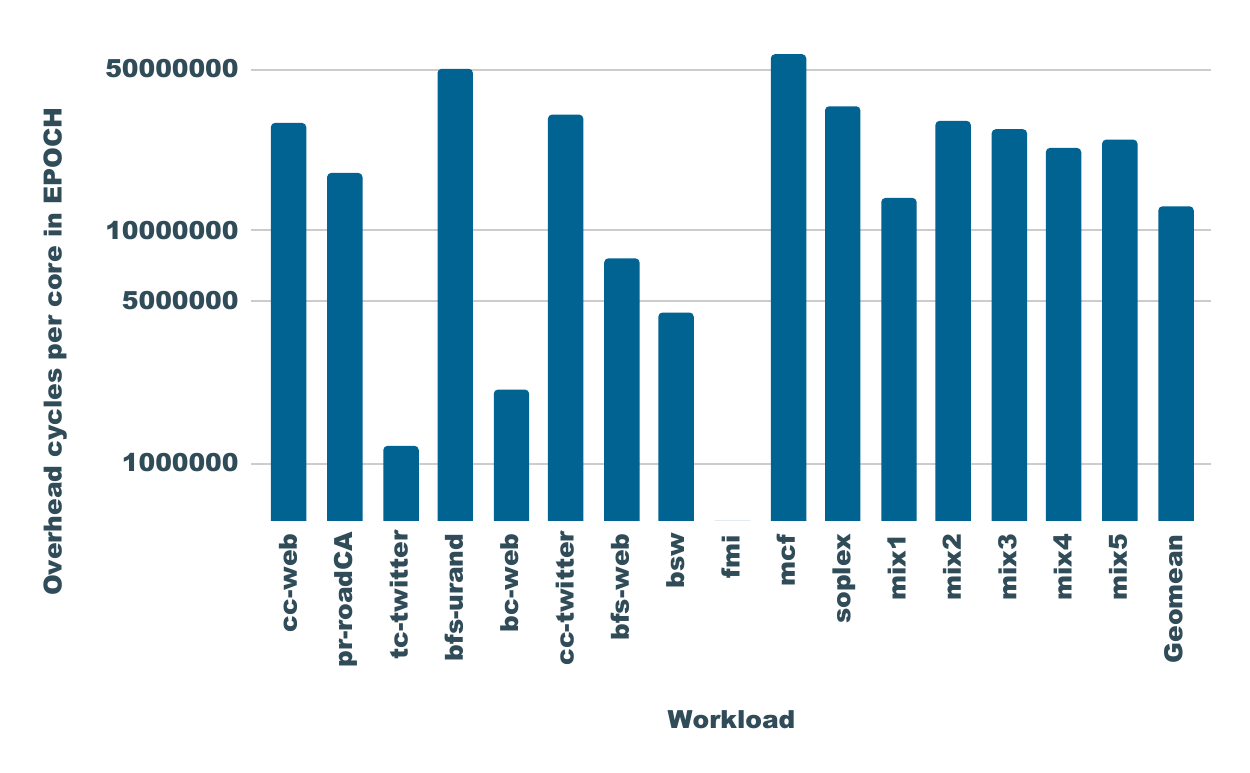}
        \caption{Accumulated overhead cycles per core due to page migration in EPOCH.}
        \label{hma_epoch_oh_cycles}
    \end{subfigure}
    \caption{Accumulated overhead cycles per core in ONFLY and EPOCH. Y-axis is in logarithmic scale.}
    \label{fig:mainfigure}
\end{figure}

To better understand the impact of page migration overheads, we study the overhead cycles of two hardware-centric approaches, on-the-fly page migration (ONFLY) \cite{OnTheFly} and epoch-based page migration (EPOCH) \cite{HMA-HS}, the detailed experimental setup is discussed in Section \ref{exp_setup}. Fig.~\ref{hma_onfly_oh_cycles_due_to_ar} and  \ref{hma_epoch_oh_cycles} displayed in logarithmic scale, show that on an average, 12641913 and 12775349 overhead cycles per core for ONFLY and EPOCH respectively, are spent for page migration in heterogeneous memory system. In EPOCH, the overhead is more prominent due to periodic nature of migration of a number of pages at the same time, leading to a greater overhead as shown in Fig.~\ref{hma_epoch_oh_cycles}. However, this overhead is relatively less in more dynamic and real-time approach such as ONFLY as shown in Fig.~\ref{hma_onfly_oh_cycles_due_to_ar}. The key objective of our work is to identify opportunities for reducing these cycles and improving overall system efficiency.       



\begin{figure}[htp]
    \centering
    \begin{subfigure}{0.48\textwidth}
        \centering
        \includegraphics[width=0.8\linewidth]{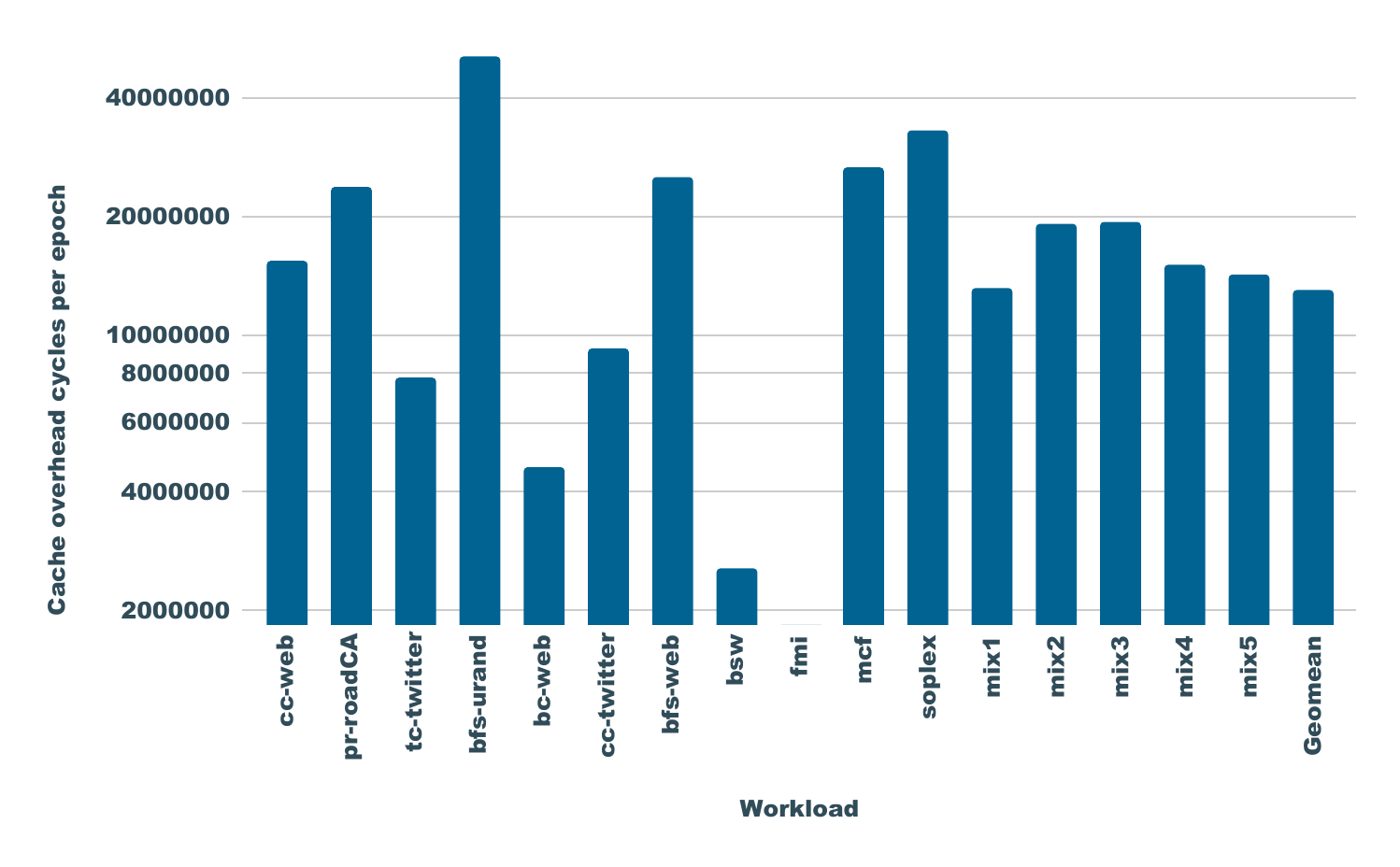}
         \caption{Cache overhead cycles per epoch.}
        \label{cache_oh_per_epoch}
    \end{subfigure}
    \hfill 
    \begin{subfigure}{0.48\textwidth}
        \centering
        \includegraphics[width=0.8\linewidth]{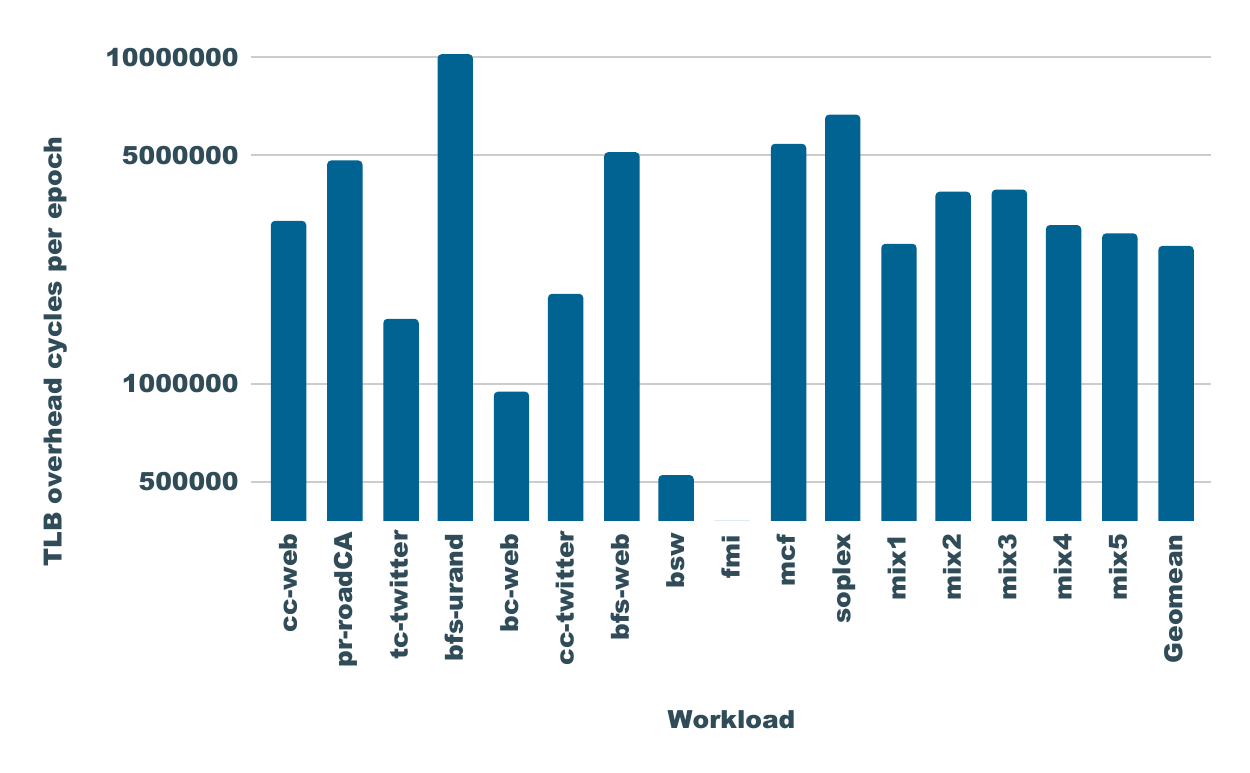}
        \caption{TLB overhead cycles per epoch.}
        \label{tlb_oh_per_epoch}
    \end{subfigure}
    \caption{Cache and TLB overhead cycles per epoch in EPOCH. Y-axis is in logarithmic scale.}
    \label{fig:mainfigure}
    \vspace{-3mm}
\end{figure}

The current approach to memory page relocation presents several inefficiencies, most notably the overhead associated with frequent TLB invalidations and cache line updates, termed as Address Reconciliation \cite{OnTheFly}. For instance, Fig.~\ref{cache_oh_per_epoch} and \ref{tlb_oh_per_epoch} displayed in logarithmic scale, show overhead cycles per epoch due to cache line invalidations and TLB shootdown in EPOCH technique, showcasing that 13032887 and 2656159 cycles respectively are consumed as part of address reconciliation overhead. This inefficiency becomes a bottleneck in scenarios requiring high-performance computation or real-time data processing. The need for a novel solution becomes evident when considering the growing reliance on complex memory architectures in emerging technologies. 

To this end, we propose a novel hardware-based solution called {\tt Duon}, that eliminates the need for frequent TLB shootdowns and cache line invalidations due to page migration by storing the updated mapping information directly within the TLB and the page table. By addressing the limitations of current OS-based solutions and harnessing the capabilities of hardware, our approach aims to unlock the full potential of heterogeneous memory systems, paving the way for more efficient and scalable memory architectures.

\section{Our Proposal: {\tt Duon}}
\label{our_idea}
To completely eliminate the invalidation operations, we propose {\tt Duon} that extends the traditional TLB and page table to include new remapped physical addresses. For current illustration of {\tt Duon}, we use a threshold-based page migration policy unless stated otherwise, wherein a page is categorized as hot, if it crosses a certain predetermined threshold. However, unlike epoch-based policy, where pages are transferred after an epoch length, in our policy, the pages are transferred as soon as they become hot \cite{OnTheFly}. It can be noted that, our proposal {\tt Duon}, can work with any underlying page migration policy.

In {\tt Duon}, the initial mapping of a virtual page (for example, $VA_1$), to its initial unified physical address which is visible to the OS, say $UA_1$, is preserved even when the page is migrated to a new remapped physical address, say, $RA_1$. The migration policies would track the hotness of pages using $UA$ in \texttt{Duon}. During the migration process, an {\tt Ongoing Migration Flag} is set to `1' to indicate that the migration of the page is in progress. Once the migration is complete, another flag, namely {\tt Migrated Flag}, is set to `1'. When the virtual-to-physical address mapping for a page is accessed, the system checks the {\tt Migrated Flag}. If this flag is set to `1', the page is accessed from its new physical address ($RA$); otherwise, it is accessed from its initial physical address ($UA$). 

\begin{figure}[h]
  \centering

  \begin{subfigure}[t]{\textwidth}
    \centering
    \resizebox{\textwidth}{!}{
  \begin{tabular}
  {|c|c|c|c|c|c|c|c|c|}

    \hline 
    \cellcolor{gray!10} \makecell{\textbf{Virtual} \\ \textbf{Address}} & 
    \cellcolor{gray!10} \makecell{\textbf{Initial } \\
    \textbf{Physical} \\
    \textbf{Address}} & 
    \cellcolor{gray!10} \textbf{Valid Bit} &
    \cellcolor{gray!10} \textbf{Dirty Bit} &
    \cellcolor{yellow!20} \makecell{\textbf{Remapped } \\ 
    \textbf{Physical} \\ 
    \textbf{Address}} & 
    \cellcolor{yellow!20} \makecell{\textbf{Migrated Flag} \\ \textbf{(Initial(0) / } \\
    \textbf{Remapped(1))} } & 
    \cellcolor{yellow!20} \makecell{\textbf{Ongoing } \\ 
    \textbf{Migration} \\ 
    \textbf{Flag}} & 
    \cellcolor{yellow!20} \textbf{Pair} & 
    \cellcolor{yellow!20} \makecell{\textbf{Buffer } \\
    \textbf{Residency Flag} \\ 
    \textbf{(Hot(1)/Cold(0))}} \\
    \hline $VA_1$ & $UA_1$ & 1 & 0 & $RA_1$ & 1 & 0 & 1 & 1  \\
    \hline $VA_2$ & $UA_2$ & 1 & 0 & $RA_2$ & 1 & 0 & 0 & 0  \\
    
    \hline
  \end{tabular}
  }
    
    \caption{Extended Page Table (EPT) Structure.}
    \label{ext_pt_struct}
  \end{subfigure}
  \vspace{1em} 

    \begin{subfigure}[t]{\textwidth}
    \centering
\resizebox{0.85\textwidth}{!}{
  \begin{tabular}
    {|>{\columncolor{gray!20}}c|
    >{\columncolor{gray!20}}c|
    >{\columncolor{gray!20}}c|
    >{\columncolor{gray!20}}c|
    >{\columncolor{yellow!20}}c|
    >{\columncolor{yellow!20}}c|
    >{\columncolor{yellow!20}}c|}
    
    \hline 
    \makecell{\textbf{Virtual Address}} & 
    \makecell{\textbf{Initial Physical} \\
    \textbf{Address}} &
    \makecell{\textbf{Valid Bit}} &
    \makecell{\textbf{Dirty Bit}} &
    \makecell{\textbf{Remapped Physical} \\ 
    \textbf{Address}} & 
    \makecell{\textbf{Migrated Flag} \\ \textbf{(Initial(0) / Remapped(1))} } &
    \makecell{\textbf{Ongoing } \\ 
    \textbf{Migration} \\ 
    \textbf{Flag}} \\
    
    \hline
    \end{tabular}
  }
  
  \caption{Extended TLB Structure.}
  \label{ext_tlb_struct}
  \end{subfigure}
  
  \caption{Extended Page Table and TLB Structure. (Note: Column headers shaded in gray represent existing architectural fields, while those highlighted in yellow denote new extensions used in {\tt Duon}.) 
  }
  \label{EPT_structure}
\end{figure}

In {\tt Duon}, the central structure is the Extended Page Table (EPT) which not only includes the virtual-to-physical address mappings, {\tt Migrated Flag} and {\tt Ongoing Migration Flag}  but also stores several key metadata flags to effectively manage and track the migration process. The {\tt Pair Flag} indicates whether the migration is a paired migration (when pages are swapped between locations) or a one-way migration (when a page is moved to an unallocated memory location). This work adopts the use of hot and cold buffers, wait queue, and migration queue as proposed by prior work \cite{OnTheFly}. Hot and cold buffers temporarily store data from pages during migration. Wait queue stores read and write requests from the LLC targeting pages currently undergoing migration and are served from hot/cold buffers. Migration queue in each memory controller is used to service read requests issued by migration controller for the migrating pages. The {\tt Buffer Residency Flag} in EPT specifies whether the page being migrated is currently stored in the hot buffer (frequently accessed data) or the cold buffer (less frequently accessed data). 



For updation of a new physical address and {\tt Migrated Flag} in Fig.~\ref{EPT_structure}, following steps are taken at subsequent timestamps: $VA_1 \rightarrow UA_1$ and $VA_2 \rightarrow UA_2$ are the initial virtual to physical address mapping, wherein $VA_1$ and $VA_2$ are two virtual addresses, $UA_1$ and $UA_2$ are the unified physical addresses that are visible to the OS, which means the physical address can be either from fast or slow memory in unified (flat) address space. $RA_1$ and $RA_2$ represent the actual location of a page after migration in physical memory, i.e., fast and slow memory, whose page migration process is shown in the Fig.~\ref{HMA_EPT_Block_Diagram}, wherein shaded portion show extended storage for keeping migration related metadata. 

\par 
Fig.~\ref{HMA_EPT_Block_Diagram} shows 16 cores in a multi-core system. Each core has a TLB that caches recent virtual-to-physical address translations, and private cache to store frequently accessed data. A larger shared cache is shared among all the cores. On-chip migration controller contains hot and cold buffers along with page migration logic. Hot/cold buffers temporarily store data for pages under migration.  {\tt Bit Vector} stores the migration status of individual cache lines within a page. If a bit in the vector is set to `1', it indicates that the corresponding cache line has completed migration, and any new memory requests for that line are redirected to the new physical address ($RA_1$) with an added line offset. Conversely, if the bit is `0', memory requests are served from the hot or cold buffer, as determined by the {\tt Buffer Residency Flag}, shown in Fig.~\ref{EPT_structure}. Once the migration is complete, the {\tt Bit Vector} is reset. Migration controller maintains a wait queue that temporarily stores read and write requests from the LLC targeting pages in the process of migration. These requests are typically handled by hot and cold buffers. HBM and PCM/DDR4 are the faster and slower memories that constitute hybrid memory system.


\subsection{Case 1: Initial Page Migration and Remapping Procedure}

\begin{figure}[tb]
  \centering
  \includegraphics[width=0.8\linewidth,height=0.8\linewidth,keepaspectratio]{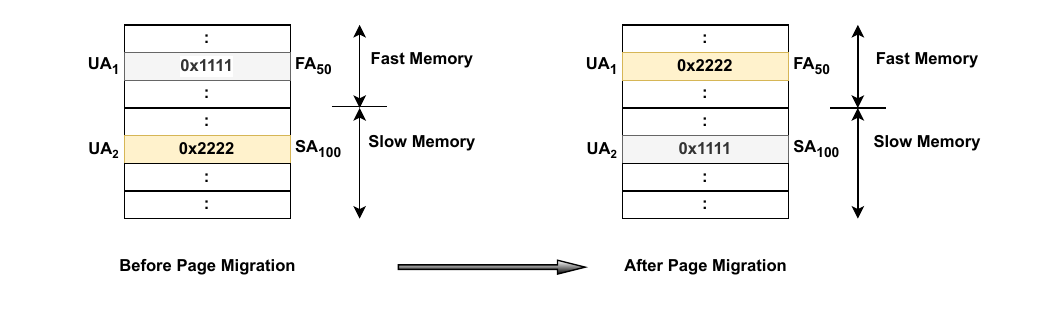}
  \caption{Page migration demonstration in Unified Address Space.}
  \label{unified_addr_space}
  \Description{Fig. showing page migration demonstration in Unified Address Space.}
  \vspace{-4mm}
\end{figure}

Consider a case wherein we have two pages, one in faster and another in slower memory, having virtual addresses $VA_1$ and $VA_2$, that are mapped to unified addresses $UA_1$ and $UA_2$ respectively in a flat address space as shown in Fig.~\ref{unified_addr_space}. These unified addresses are further mapped to physical addresses in either fast or slow memory. $UA_1 \rightarrow FA_{50}$ is located in fast memory and contains data (say) \texttt{0x1111}, whereas $UA_2 \rightarrow SA_{100}$ is located in slow memory and contains data (say) \texttt{0x2222} before page migration.
In conventional page migration without using unified address mapping, virtual addresses are directly mapped to actual physical addresses as shown in following scenario:

\textbf{Before migration:} \\
\hspace*{1cm} $VA_1 \rightarrow FA_{50}$ (0x1111) \\
\hspace*{1cm} $VA_2 \rightarrow SA_{100}$ (0x2222) \\

\textbf{After migration:} \\
\hspace*{1cm} $VA_1  \rightarrow SA_{100}$ (0x1111) \\
\hspace*{1cm} $VA_2   \rightarrow FA_{50}$ (0x2222) \\


Since the physical addresses associated with each virtual address have changed after page migration, the system must invalidate old TLB entries and insert new ones, flush or invalidate cache lines that may contain stale data. This process of TLB shootdown and cache line invalidation introduces significant overhead (discussed in Section~\ref{motive}) especially when there are frequent page migrations. \par 

The unified address layer between virtual and actual physical address helps to eliminate this overhead. Each unified address (UA) maps to a physical address, and the virtual-to-unified address mapping remains unchanged during migration. Now, instead of updating TLB and caches, we store two physical addresses per virtual address. 
Thus, the effective memory access paths before migration become: \\ 
\hspace*{1cm} $VA_1 \rightarrow UA_1 \rightarrow FA_{50}$ (0x1111) \\
\hspace*{1cm} $VA_2 \rightarrow UA_2 \rightarrow SA_{100}$ (0x2222) \\

To optimize performance, let us consider that the migration controller initiates page migration, swapping the physical locations of the pages associated with $UA_1$ and $UA_2$. After page migration, $UA_1 \rightarrow FA_{50}$ now contains 0x2222 and $UA_2 \rightarrow SA_{100}$ now contains 0x1111. The updated access paths become: \\
\hspace*{1cm} $VA_1 \rightarrow UA_1 \rightarrow SA_{100}$ (0x1111) \\
\hspace*{1cm} $VA_2 \rightarrow UA_2 \rightarrow FA_{50}$ (0x2222) \\

\begin{figure}[htbp]
  \centering
  \includegraphics[width=0.6\linewidth,height=0.6\linewidth,keepaspectratio]{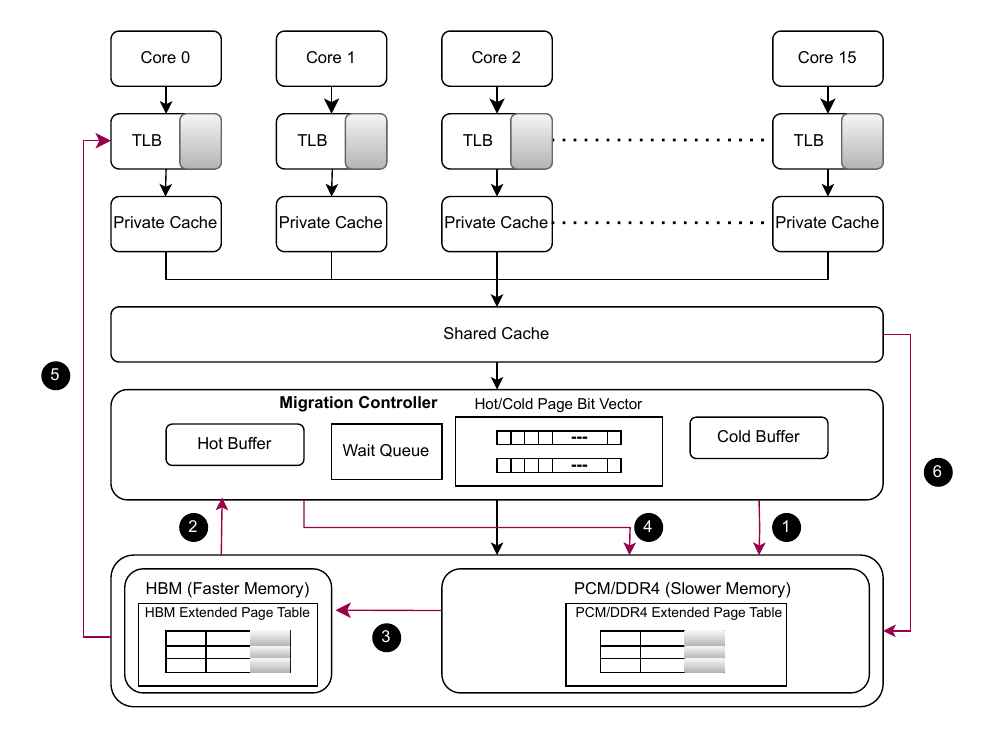}
  \caption{Overview of steps involved in Page Migration. }
  \label{HMA_EPT_Block_Diagram}
  \Description{A block diagram showing the overview of page migration steps.}
\end{figure}

This migration is transparent to the OS, as the virtual-to-unified address mapping does not change, and the TLB and caches still observe the same unified physical address even after page migration. The step-by-step migration process is illustrated in Fig.~\ref{HMA_EPT_Block_Diagram} and the corresponding values after each step are summarized in Table \ref{step_wise_page_migration}.

\begin{table}[htbp]
  \centering
  \scriptsize
  \caption{Step-wise page migration process. Column header shown in blue color are included solely for explanation purposes and is not part of the actual metadata.}
  \label{step_wise_page_migration}
  \resizebox{\textwidth}{!}{
  \begin{tabular}
  {|c|c|>{\centering\arraybackslash}p{1.5cm}|c|c|c|c|c|c|c|}
    
    \hline 
    \cellcolor{blue!20} \textbf{Step No.} &
    
    \cellcolor{gray!20} \makecell{\textbf{Virtual} \\ \textbf{Address}} & 
    \cellcolor{gray!20} \makecell{\textbf{Initial } \\
    \textbf{Physical} \\
    \textbf{Address}} & 
    \cellcolor{yellow!20} \makecell{\textbf{Remapped } \\ 
    \textbf{Physical} \\ 
    \textbf{Address}} & 
    \cellcolor{blue!20} \textbf{Content} &
    \cellcolor{blue!20} \textbf{Status} &
    \cellcolor{yellow!20} \makecell{\textbf{Migrated Flag} \\ \textbf{(Initial(0) / } \\
    \textbf{Remapped(1))} } & 
    \cellcolor{yellow!20} \makecell{\textbf{Ongoing } \\ 
    \textbf{Migration} \\ 
    \textbf{Flag}} & 
    \cellcolor{yellow!20} \textbf{Pair} & 
    \cellcolor{yellow!20} \makecell{\textbf{Buffer } \\
    \textbf{Residency Flag} \\ 
    \textbf{(Hot(1)/Cold(0))}} \\
    \hline 1 & $VA_1$ & $UA_1 (FA_{50})$ & - & 0x1111 & - & 0 & 0 & 0 & - \\
    & $VA_2$ & $UA_2 (SA_{100})$ & - & 0x2222 & - & 0 & 0 & 0 & - \\
    \hline 2 & $VA_1$ & $UA_1$ & - & - & On hold & 0 & 1 & 1 & 1 \\
    & $VA_2$ & $UA_2$ & - & 0x2222 & - & 0 & 0 & 0 & 0 \\
    \hline 3 & $VA_1$ & $UA_1$ & - & 0x2222 & On hold & 0 & 1 & 1 & 1 \\
    & $VA_2$ & $UA_2$ & - & 0x2222 & - & 0 & 0 & 0 & 0 \\
    \hline 4 & $VA_1$ & $UA_1$ & $SA_{100}$ & 0x1111 & On hold & 0 & 1 & 1 & 1 \\
    & $VA_2$ & $UA_2$ & $FA_{50}$ & 0x2222 & - & 0 & 0 & 0 & 0 \\
    \hline 5 & $VA_1$ & $UA_1$ & $SA_{100}$ & 0x1111 & - & 1 & 0 & 1 & 0 \\
    & $VA_2$ & $UA_2$ & $FA_{50}$ & 0x2222 & - & 1 & 0 & 1 & 0 \\
    
    \hline
  \end{tabular}
  }
\end{table}

Step \filledcircled{1} in Fig.~\ref{HMA_EPT_Block_Diagram} shows decision to move hot page with physical address $SA_{100}$ into fast memory: At this point, $VA_1 \rightarrow UA_1 \rightarrow FA_{50}$ (0x1111) and $VA_2 \rightarrow UA_2 \rightarrow SA_{100}$ (0x2222) are the initial virtual-to-physical address mappings in the fast and slow memory respectively. On-chip Migration Controller takes the decision to move 0x2222 from the slower to the faster memory. If there is no free page available in the faster memory, $FA_{50}$ (0x1111) (say) is selected as victim page to move to the slower memory. 

\filledcircled{2} Move 0x1111$\,($at  $FA_{50})$ into hot buffer: In this step, the incoming requests to $VA_1 \rightarrow UA_1$ are put on hold, i.e., stall the $UA_1$'s requests. However, $UA_2$'s requests continue to be served from slow memory. At this stage, the status of pages under migration is shown in Step 2 in Table \ref{step_wise_page_migration}. This step concludes when the movement of 0x1111 to hot buffer completes.

\filledcircled{3} Move a copy of 0x2222$\,($at $SA_{100})$ into fast memory: At this point, $UA_1$'s requests continue to stall and $UA_2$'s requests are served from slow memory. This step concludes when writing the content of $UA_2$, i.e. 0x2222 to location $UA_1$ completes. The status is given by Step 3 of Table~\ref{step_wise_page_migration}.

\filledcircled{4} Move 0x1111 from buffer to slow memory: At this point, $UA_2$'s request is served from fast memory as 0x2222 is moved into fast memory in Step \filledcircled{3}. But $UA_1$'s requests continue to stall, with the status as shown in Step 4 of Table \ref{step_wise_page_migration}. 
This step concludes when writing the data of 0x1111 from the hot buffer to location $SA_{100}$ completes.

\filledcircled{5} Page migration complete: At this point, $UA_2$'s requests are served from fast memory and $UA_1$'s requests from slow memory. The mapping at this stage is shown in Step 5 of Table \ref{step_wise_page_migration}. 
This step concludes when all the corresponding entries in the EPT are updated.

For data consistency, we propose that the TLB entries also need to be extended to store the remapped physical address, migrated flag, ongoing migration flag along with the initial physical address (Fig.~\ref{HMA_EPT_Block_Diagram}). Accordingly, in Step \filledcircled{5}, after the page migration is complete, the remapped physical addresses in page table are also updated in TLB if they reside. Please note that in the TLB, entries will be similar to tuples, <$VA_1$, $UA_1$, $-$, 0, 0>, and <$VA_2$, $UA_2$, $-$, 0, 0> before migration; and <$VA_1$, $UA_1$, $SA_{100}$, 1, 0>, and <$VA_2$, $UA_2$, $FA_{50}$, 1, 0> after migration. 0 and 1 here are the migrated and ongoing migration flag that comes into consideration only for a LLC miss, wherein if migrated flag is 0, initial physical address is used to access the memory, otherwise remapped physical address is used. \par

\begin{figure}[h]
  \centering
  \includegraphics[width=0.6\linewidth,height=0.6\linewidth,keepaspectratio]{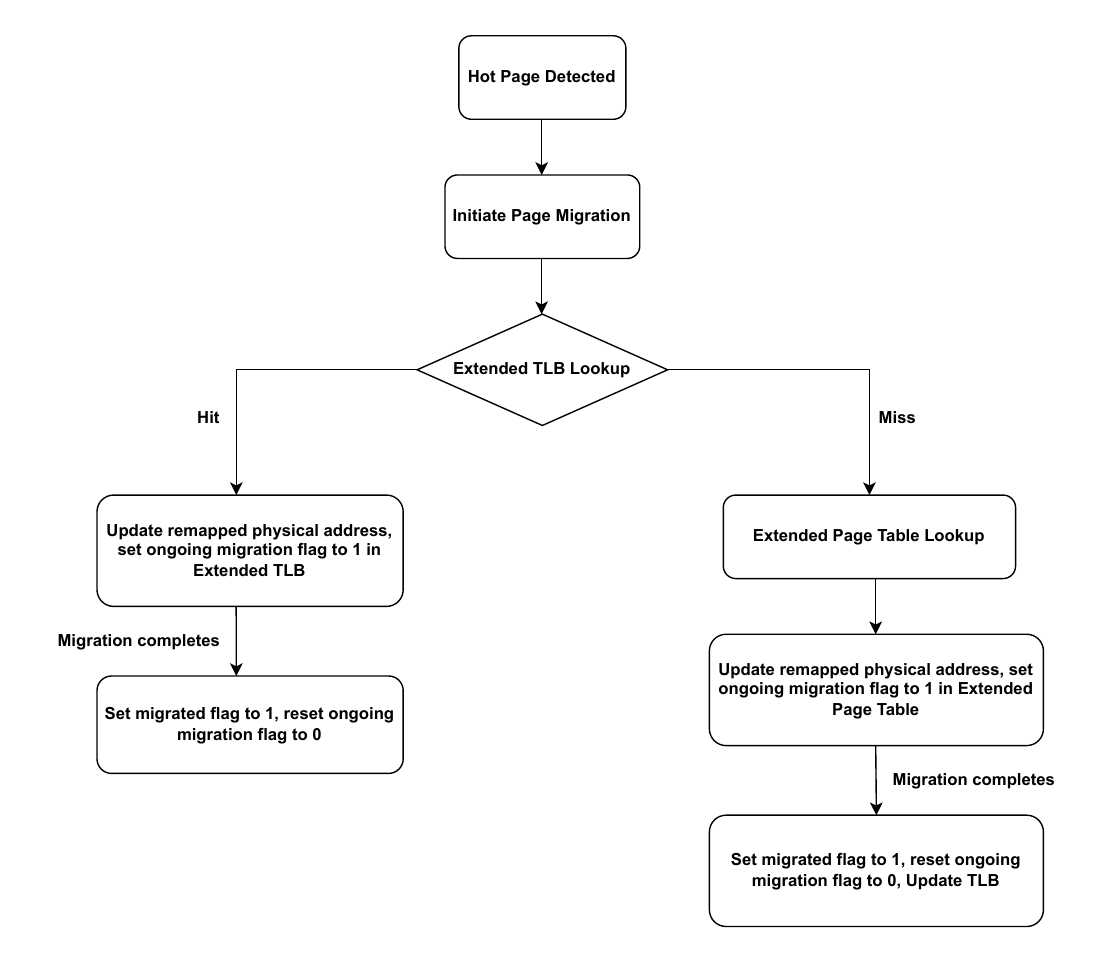}
  \caption{Remapped Physical Address updation in Extended TLB and Page Table.}
  \label{ext_tlb_update}
  \Description{A flowchart showing the Remapped Physical Address updation in Extended TLB and Page Table}
\end{figure}

Fig.~\ref{ext_tlb_update} shows the sequence of operations that occur during page migration in a memory system. The process begins when a hot page is detected, prompting the initiation of page migration. At this point, extended TLB lookup happens. If the lookup results in a hit, the remapped physical address is updated directly in the extended TLB, and the ongoing migration flag is set to 1. If the lookup misses, remapped physical address in extended page table is updated, setting the ongoing migration flag to 1. Once the page migration completes, both the extended TLB and page table update their respective entries by setting the migrated flag to 1 and resetting the ongoing migration flag to 0. Additionally, the TLB is updated to reflect the new mapping.

Step \filledcircled{6} in the Fig.~\ref{HMA_EPT_Block_Diagram} shows the Last Level Cache (LLC) or Shared Cache miss requests that are served from main memory after page migration is complete. This occurs following an extended TLB/page table lookup, that determines whether the requested page has been migrated or not. It is pertinent to note that during normal execution, the standard TLB entry is accessed to perform virtual-to-physical address translation. However, in the event of an LLC miss, there is a second access to the extended TLB/page table to determine correct physical address based on the migration status of the data.

\begin{table}[h]
  \centering
  \scriptsize
  \caption{Changes to the TLB and Cache look up process.}
  \label{TLB_Cache_lookup}
  \begin{tabular}
  {|c|c|l|}
  \hline
  \textbf{TLB}  & \textbf{Shared Cache} & \textbf{Description} \\ 
 \textbf{ look-up} & \textbf{look-up} & \\ \hline
  Hit & Hit & UA provided by TLB is used in Cache.\\ \hline
  Hit & Miss & Upon a Cache miss, if the page is noted to be migrated, the TLB\\ 
  & & provides RA to the memory. If the page is not migrated, UA can be\\
  & &  used directly for memory look-up. When the data returns to the\\
  & &  Cache, UA is used for Tag. \\ \hline
  Miss & Hit & TLB forwards the request to EPT. EPT responds with both UA and RA\\ 
  & &  (in case of migration) to the TLB. TLB provides UA to the Cache for\\
  & &  look-up which resulted in a hit.\\ \hline
  Miss & Miss & TLB forwards the request to EPT. EPT responds with both UA and RA\\ 
  & &  (in case of migration) to the TLB. TLB provides UA to the Cache for\\
  & & look-up which resulted in a miss. RA is used in the memory for look \\
  & & -up if the page is migrated, otherwise UA is used. \\ \hline
  
\hline
  \end{tabular}
\end{table}

The tag entries in the on-chip caches will continue to work to be valid, as <$VA_1$, $UA_1$> and <$VA_2$, $UA_2$> mapping will still hold. Such a design needs subtle but important changes to the working of TLB and caches. Table~\ref{TLB_Cache_lookup} outlines the interaction between TLB and shared cache during memory access operations, focusing on whether each component registers a hit or miss and its response in each scenario. When both the TLB and shared cache register a hit, the process is straightforward. TLB provides the unified address (UA), that is directly used by the cache to retrieve the requested data. This represents the most efficient path, with no need for further address translation or memory access. When TLB registers a hit but the cache registers a miss, it has to be determined whether the page has been migrated. If migration has occurred, TLB provides remapped physical address (RA) to access the data from memory. If the page has not been migrated, the UA can still be used for memory access. Once the data is retrieved from memory and placed into the cache, the UA is used as the tag for future lookups. \par 
When TLB registers a miss but cache registers a hit, TLB forwards the request to EPT. EPT responds with both UA and RA (if migration is involved). TLB then provides UA to perform a cache lookup, which in this case results in a hit, indicating that the data was already present in the cache despite the initial TLB miss. Lastly, when both TLB and cache register a miss, TLB again consults the EPT. EPT provides UA and RA, and the TLB uses UA to attempt a cache lookup, that fails. At this point, system must access memory. If the page has been migrated, RA is used, otherwise UA is used to access memory. This is most resource-intensive path, involving both address translation and memory access. Thus, TLB and cache can continue to work with UA even after migration, without the need for address reconciliation. This way, {\tt Duon} makes the working of TLB and on-chip caches agnostic of the page migrations.  \par

\begin{figure}[tb]
  \centering
  \includegraphics[width=0.7\linewidth,height=0.7\linewidth,keepaspectratio]{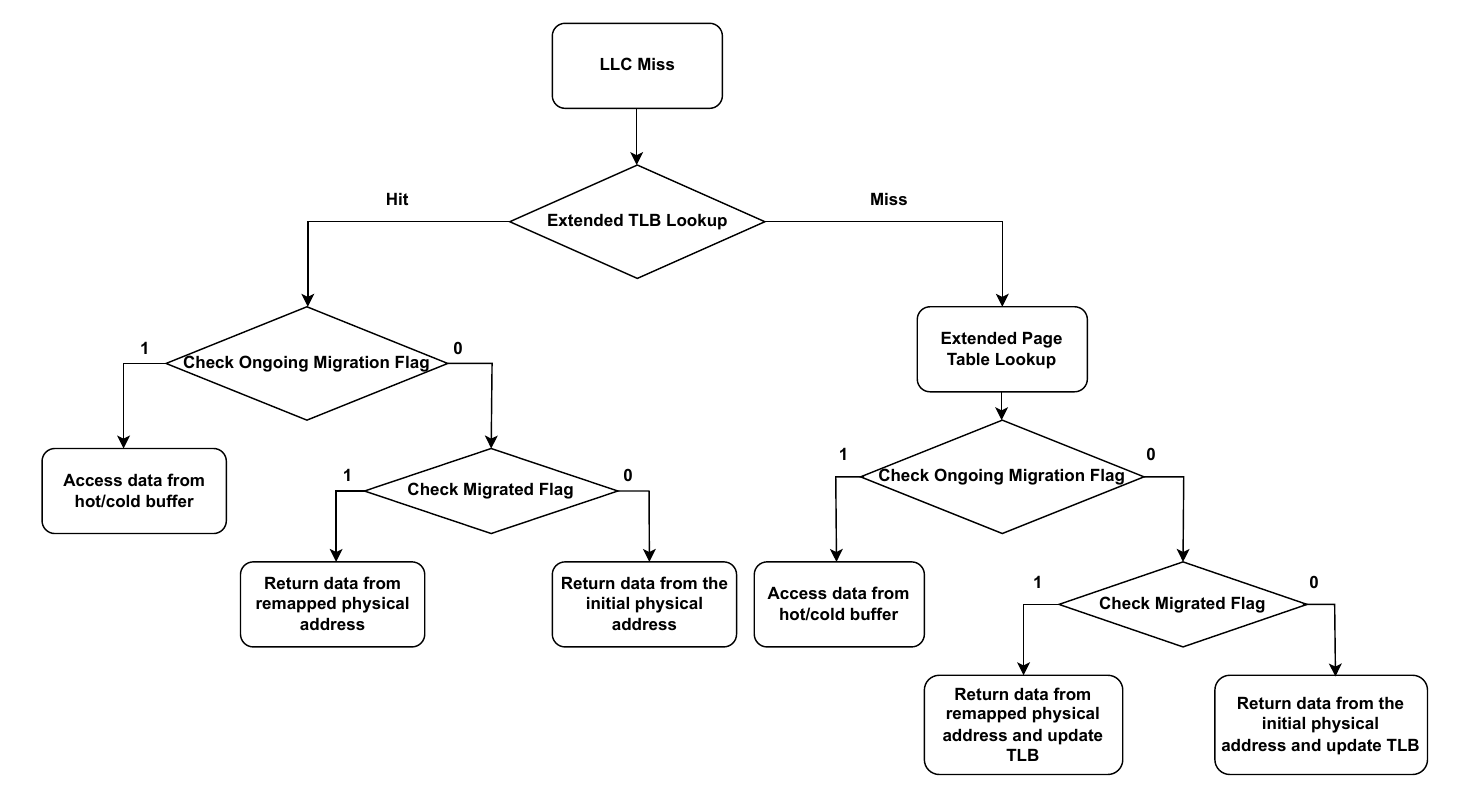}
  \caption{Extended TLB and Page Table Lookup in {\tt Duon}.} 
  \label{ext_tlb_lookup}
  \Description{A flowchart showing Extended TLB Lookup.}
\end{figure}

Fig.~\ref{ext_tlb_lookup} outlines the decision-making process when an LLC miss occurs, triggering a deeper lookup through extended TLB and page table. Initially, extended TLB lookup happens, and the flow then evaluates two flags, one indicating whether migration is ongoing, and another confirming if the page has already been migrated. Based upon these flags, decision is made whether to read data from hot/cold buffer or to use initial or remapped physical address. If page migration is ongoing, data is retrieved from hot/cold buffer, but if the page is migrated, remapped physical address is used to access data. If extended TLB lookup is a miss, extended page table lookup happens, and ongoing migration flag, and migrated flag of extended page table are checked for data access location. Upon completion of extended page table lookup, the respective page table entry is updated in extended TLB.

In case of a page fault, the working of {\tt Duon} is similar to that of the conventional page table. The OS will initiate the page request to the storage and in the mean time identifies a victim page in the memory. The entries corresponding to the victim page are marked invalid in EPT and TLB, and the cache. The new page is written to the memory (in the location of the victim page) and overwrites the entry in the EPT. Further, the working of {\tt Duon} is not tightly bound to a specific migration policy.  Implying that, the monitoring of pages usage in the fast and slow memories, identifying a page to migrate, identifying the victim page in the target memory, etc., are all done by the migration controller. The migration controller makes sure that the actual migration of the pages happens and updates its fields from time to time during the migration in TLB and EPT. Hence, {\tt Duon} can be integrated into any of the existing page migration policies such as ONFLY\cite{OnTheFly}, EPOCH \cite{HMA-HS} and ADAPT-THOLD \cite{DynAdapThold}. 


\subsection*{TLB Coherence}


{\tt Duon} as explained above stores additional metadata. With this metadata, memory accesses, especially those that miss in the LLC, can be efficiently redirected. Based on the extended TLB and page table entries, whether the page in question has been migrated is determined, and the appropriate physical address is accessed accordingly. In traditional systems, any change in the physical address mapping typically requires operating system intervention to maintain TLB coherence across all cores, often through TLB shootdowns. However, in {\tt Duon}, the initial physical address remains unchanged. Instead, the mapping from the initial to the remapped (migrated) address is updated—thus, the coherence must be preserved not by modifying the base address, but by synchronizing the new mapping across all cores.
To achieve this, we introduce a hardware-based TLB Coherence Module (TCM). This module ensures consistency of TLB entries across cores for remapped physical addresses. It is activated when the migration controller detects a hot page and initiates migration. Once triggered, TCM manages coherence transparently and efficiently without requiring software-level shootdowns. \par 


 The working of TCM to maintain consistency across multiple cores during page migration begins when the migration controller identifies a hot page and initiates its migration. To ensure that all cores are aware of the change, a signal is sent to TCM. TCM then broadcasts an update command to every core's private TLBs. Each private TLB checks whether it contains an entry corresponding to the initial physical address (UA). If such an entry is found, it is updated to set the ongoing migration flag. If no matching entry exists, no action is taken. Once the migration completes, the initiating core will send out another broadcast request with the remapped address. The other cores will check and update their private TLBs if needed. The updation involves setting migrated flag to 1, resetting ongoing migration flag to 0, and updating the remapped physical address. Once all updates are complete, TCM sends an acknowledgment back to the migration controller, confirming that coherence has been maintained across the system. This mechanism ensures that all cores operate with the correct and current memory mappings, preventing stale or incorrect data access during and after page migration.

\subsection*{How multi-core systems with shared LLC use {\tt Duon}?} 
Page migration typically introduces cache inconsistencies, particularly due to changes in physical address mappings that necessitate cache line invalidations. To address this, {\tt Duon} incorporates an indirection mechanism, where the initial physical address ($UA$) serves as an access pointer, but the actual data resides at a remapped address ($RA$) post-migration. Despite the data being migrated, caches continue to index and access content using $UA$, preserving coherence from the cache’s perspective. Upon a cache miss, the migrated flag triggers a lookup to the remapped physical address ($RA$), ensuring accurate data retrieval. Furthermore, any cache-level updates, whether in write-through or write-back policies, are directed to $RA$ via the indirection logic. This guarantees that changes are consistently propagated to the correct location in memory without relying on system-wide invalidations. Through this mechanism, \texttt{Duon} maintains cache coherence and minimizes migration-related overheads such as TLB shootdowns and cache flushing.

\subsection{Case 2: Re-migration of previously migrated page}
In cases where a previously migrated page is identified again as a hot page, {\tt Duon} supports re-migration to a new remapped address without disrupting the system’s consistency. During re-migration, the Ongoing Migration Flag is again set to 1, signaling that the page is in transition. Memory accesses consult the most recent valid remapped address, because migrated flag is set to 1 for already migrated pages during initial page migration. The Bit Vector is reused to track the migration status of individual cache lines from 
already remapped address to new remapped address. Once complete, the ongoing migration flag resets to 0, and the latest remapped address is accessed. TCM is re-activated upon re-migration trigger. It broadcasts the updated remapped address mapping to all cores holding stale TLB entries. Thus, the steps shown in Table \ref{step_wise_page_migration} are performed again for updation of existing remapped address to new remapped address upon page re-migration.

\section{Experimental Setup}
\label{exp_setup}
For analyzing {\tt Duon}, we use Ramulator \cite{Ramulator} simulator, to model and evaluate a 16-core system with a flat address memory architecture. This memory system integrates 1 GB of HBM, known for its high performance and speed; and 16 GB of PCM (16 GB DDR4 for sensitivity analysis), that provides greater capacity but has different performance characteristics. To demonstrate the adaptability of our proposal across different memory configurations, we conduct a sensitivity study using 1 GB of HBM in conjunction with 16 GB of DDR4 memory detailed in Subsection~\ref{sensitivity_analysis}. The system configuration and memory architecture are based on the simulation models developed by Shashank et al. ~\cite{OnTheFly}. By combining faster and the slower memory in a flat address space design, our setup aims to balance both performance and capacity, making it suitable for the evaluation of modern data-intensive workloads.

\begin{table}[h]
    \centering
    \scriptsize
    \caption{Baseline configuration.}
    \label{baseline_config}
    \begin{tabular}{|>{\raggedright\arraybackslash}p{2cm}|>{\raggedright\arraybackslash}p{9cm}|}
        \hline
        \textbf{Component} & \textbf{Parameter Values} \\ \hline
        Core & 16 cores, Frequency = $3.2 GHz$ \\ \hline
        L1-D Cache      & 32KB size, 64B cache block size, 4-way Set-associativity,
        data access time of 2 cycles      \\ \hline
        L2 Cache     & 16MB size, 64B cache block size, 16-way Set-associativity,
        data access time of 21 cycles     \\ \hline
        & HBM, tCAS-tRCD-tRP-tRAS: 14ns-14ns-14ns-34ns \\
  Main Memory  
  & DDR4, tCAS-tRCD-tRP-tRAS: 16ns-16ns-16ns-39ns \\
  & PCM, Read/Write latency is $80ns$/$250ns$, \\
  & 4KB page size, Epoch length is 10000$\mu$s      \\ \hline
    \end{tabular}

\end{table}

The baseline configuration of the Ramulator, that includes parameters for the memory hierarchy, processor cores is summarized in Table \ref{baseline_config}. These baseline settings are chosen to ensure a fair and accurate comparison with existing state-of-the-art memory management techniques. Additionally, our proposal requires on-chip hot and cold buffers of size 4 KB each, hot/cold page bit vector, wait queue, migration queue. The simulator allows us to analyze the performance of our proposed approach alongside other methods focusing on page migration and address reconciliation.

\begin{table}[h]
  \centering
  \scriptsize
  \caption{Benchmarks used for evaluation.}
  \label{list_benchmarks}
  \resizebox{\textwidth}{!}{
  \begin{tabular}
  {|l|l|c|}
    \hline 
    \textbf{Benchmark Suite} & \textbf{Benchmark Name} & \textbf{Memory Footprint (in GB)} \\
    \hline
    GAPBS & bc-web & 2.38 \\
          & cc-web & 6.77 \\
          & pr-roadCA & 1.04 \\
          & tc-twitter & 1.16 \\
          & cc-twitter & 7.00 \\
          & bfs-urand & 1.63 \\
          & tc-urand & 4.37 \\
          & bfs-web & 1.00 \\
    \hline      
    Genomicsbench & bsw & 3.57 \\      
                  & fmi & 6.78 \\
    \hline              
    SPEC 2006 & soplex & 1.74 \\
              & mcf & 3.05 \\
    \hline              
    PARSEC & fluidanimate & 1.04 \\

    \hline              
    MIXES & mix1 (cc-web + bc-web + bfs-web + fmi + tc-twitter + soplex + fluidanimate + bsw) x 2 & 3.06 \\
          & mix2 (bfs-urand + tc-urand + mcf + pr-roadCA + cc-twitter + bc-web + fmi + fluidanimate) x 2 & 3.41 \\
          & mix3 (fluidanimate + bsw + mcf + soplex + fmi + bfs-urand + cc-web + bc-web) x 2 & 3.37 \\
          & mix4 (tc-urand + bsw + cc-twitter + fluidanimate + bfs-web + mcf + tc-twitter + soplex) x 2 & 2.87 \\
          & mix5 (cc-web + bc-web + tc-twitter + cc-twitter + pr-roadCA + mcf + fmi + bsw) x 2 & 3.97 \\
    \hline 
  \end{tabular}
  }
\end{table}

To thoroughly assess the effectiveness of {\tt Duon}, we select high memory footprint workloads from a variety of well-established benchmark suites. These include (i) GAPBS \cite{GAPBS}, widely used for graph processing applications; (ii) Genomicsbench \cite{Genomicsbench}, focuses on computational genomics; (iii) SPEC CPU 2006 \cite{SPEC2006} benchmark suite, which offers a diverse set of workloads designed to evaluate CPU and memory performance; and (iv) PARSEC \cite{PARSEC} benchmark suite that focuses on parallel applications. The workloads along with their memory footprint are listed in Table \ref{list_benchmarks}. Our selected set of benchmarks offer sufficient diversity and coverage to effectively validate the performance, capacity, and scalability of our approach. These benchmarks provide comprehensive workloads for understanding the behavior of the proposed memory system under different real-world scenarios.
The application traces are run to their completion for each experiment. We consider No Migration as the baseline, unless stated otherwise. We study the effectiveness of {\tt Duon} by deploying it along with three state of the art hardware based page migration techniques, mainly, EPOCH, ONFLY and ADAPT-THOLD explained in Section~\ref{on-the-fly-adapt}. 

\section{Results and Analysis}
\label{results_analysis}

\begin{figure}[t]
    \centering
    \begin{subfigure}{0.48\textwidth}
        \centering
        \includegraphics[width=0.8\linewidth, keepaspectratio, height=6cm]{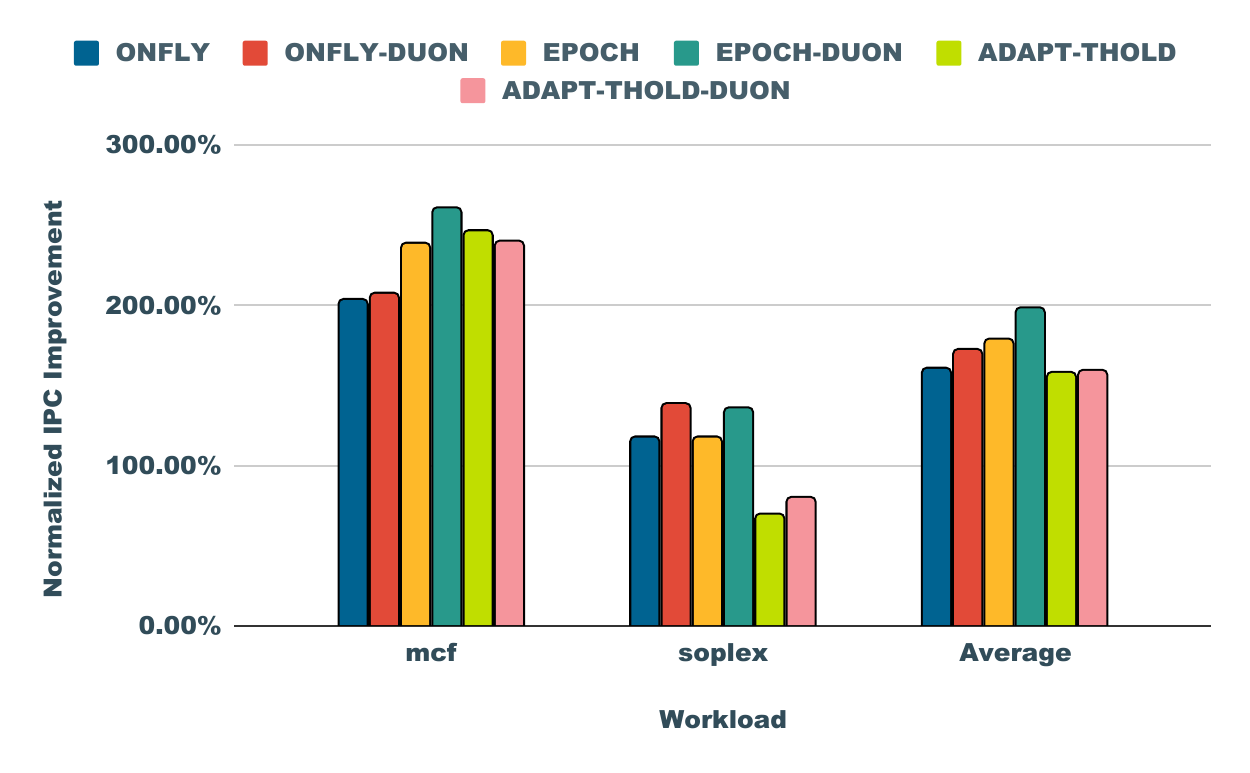}
        \caption{Normalized IPC improvemment of various HMA techniques for highly migration friendly workloads w.r.t. baseline.}
        \label{ipc_imp_over_no_mig_migr_friendly}
    \end{subfigure}
    \hfill 
    \begin{subfigure}{0.48\textwidth}
        \centering
         \includegraphics[width=0.8\linewidth, keepaspectratio, height=6cm]{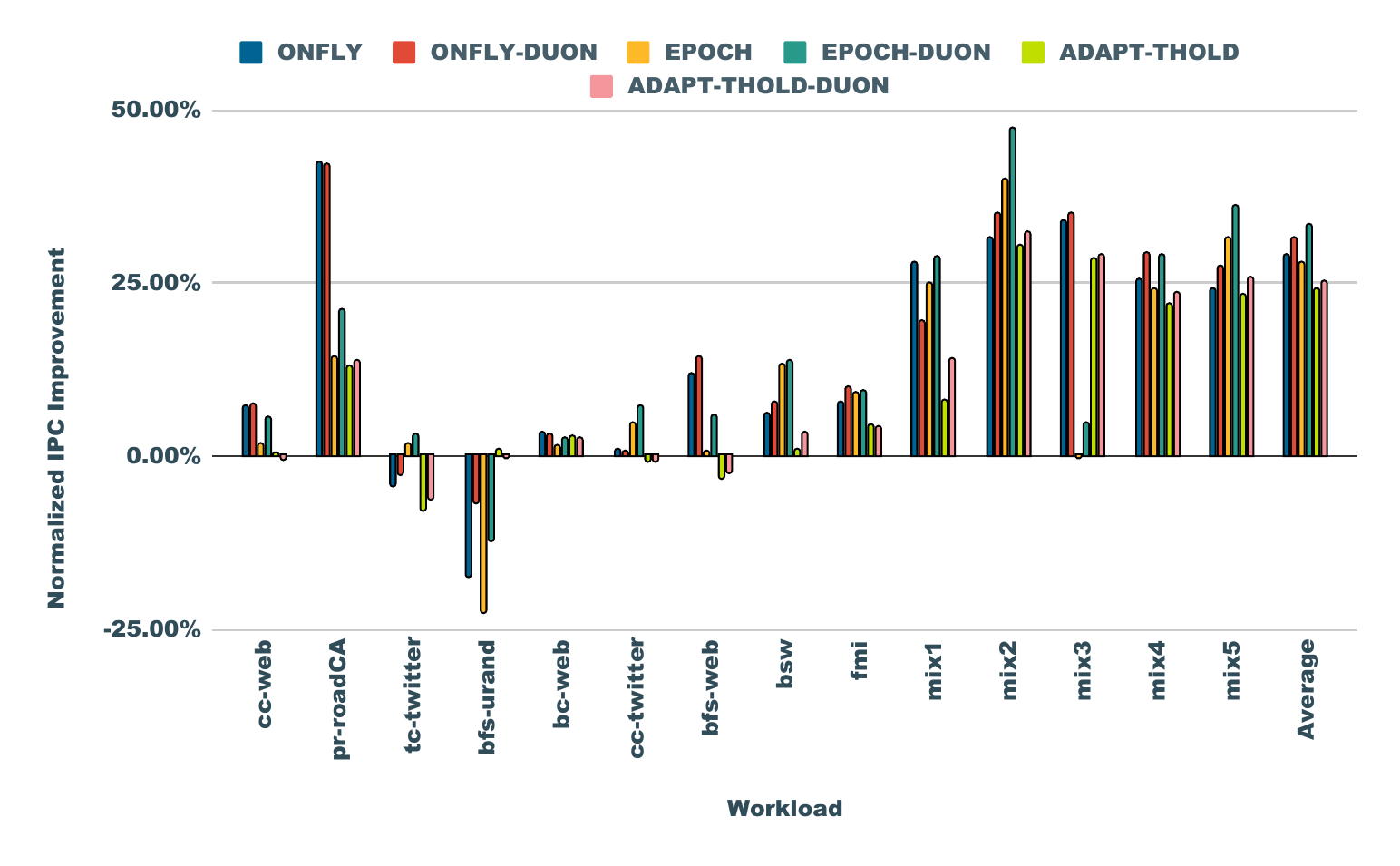}
          \caption{Normalized IPC improvemment of various HMA techniques w.r.t. baseline.}
        \label{ipc_imp_over_no_mig}
    \end{subfigure}
    \caption{Normalized IPC improvement of various HMA techniques, for threshold value of 64, HBM Size: 1 GB, PCM Size: 16 GB, w.r.t. No Migration as the baseline.}
    \label{ipc_imp_no_mig_baseline}
    \vspace{-3mm}
\end{figure}

This section analyzes the performance improvement  achieved by integrating {\tt Duon} with (i) ONFLY \cite{OnTheFly}, (ii) EPOCH \cite{HMA-HS}, and (iii) ADAPT-THOLD (ONFLY with adaptive threshold) \cite{DynAdapThold} policies, and compares the performance across the selected workloads. In Fig.~\ref{ipc_imp_over_no_mig_migr_friendly} and~\ref{ipc_imp_over_no_mig}, for ease of understanding, the results are split into two categories, for highly migration-friendly workloads (namely, {\it mcf} and {\it soplex}) and the remaining fourteen workloads (from Table~\ref{list_benchmarks}). The average IPC improvement of the six techniques for the 14 workloads in Fig.~\ref{ipc_imp_over_no_mig} highlights the effectiveness of integrating {\tt Duon}. The standalone techniques—ONFLY, EPOCH, and ADAPT-THOLD achieve an average IPC improvement of 29.00\%, 27.81\%, and 24.23\%, respectively, over the no-migration baseline. However, when {\tt Duon} is integrated, noticeable enhancements are observed: ONFLY-DUON improves to 31.49\%, EPOCH-DUON rises to 33.26\%, and ADAPT-THOLD-DUON achieves an average improvement of 25.29\%, on an average. 
While {\tt Duon} integration outperforms the other techniques overall, the results for ADAPT-THOLD-DUON indicate that {\tt Duon} integration provides only marginal improvements for a technique originally designed for adaptive threshold-based migration. This is because \texttt{Duon}'s core benefit is to reduce page migration overhead due to TLB shootdown and cache line invalidation, ADAPT-THOLD however is already designed to minimize unnecessary page migrations by tuning hotness threshold dynamically, showing lesser visible impact of \texttt{Duon} integration relative to ONFLY and EPOCH techniques. Fig.~\ref{ipc_imp_over_no_mig_migr_friendly} shows a similar trend in IPC improvement for highly migration friendly workloads, providing an average IPC improvement of 7.30\%, 11.32\%, and 1.21\% for ONFLY, EPOCH, and ADAPT-THOLD respectively with {\tt Duon} integration.



\begin{figure}[t]
    \centering
    \begin{subfigure}{0.48\textwidth}
        \centering
        \includegraphics[width=0.8\linewidth, keepaspectratio, height=6cm]{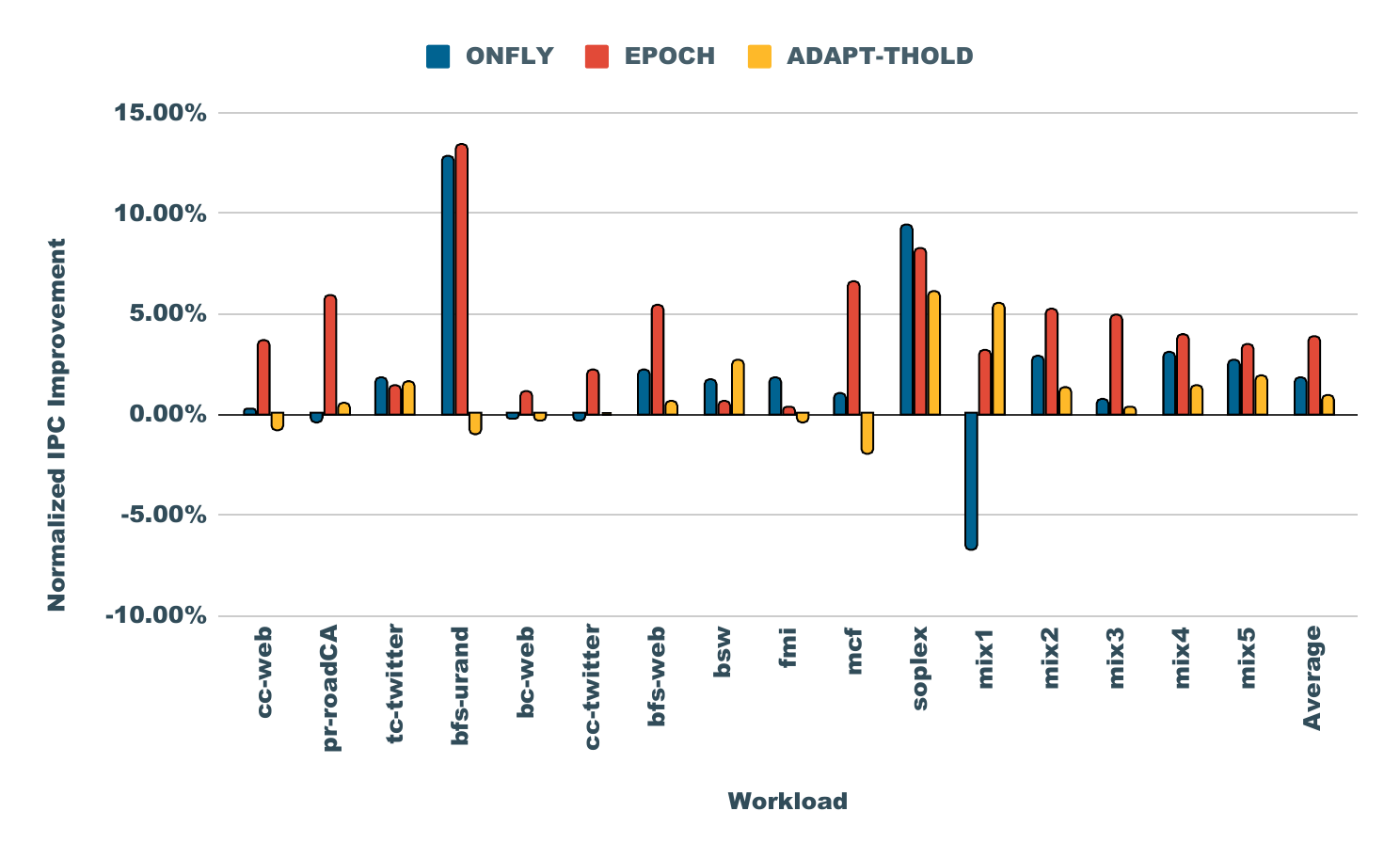}
        \caption{Normalized IPC improvement of ONFLY, EPOCH, and ADAPT-THOLD, when integrated with DUON.}
        \label{ipc_imp_with_ept}
    \end{subfigure}
    \hfill 
    \begin{subfigure}{0.48\textwidth}
        \centering
         \includegraphics[width=0.8\linewidth, keepaspectratio, height=6cm]{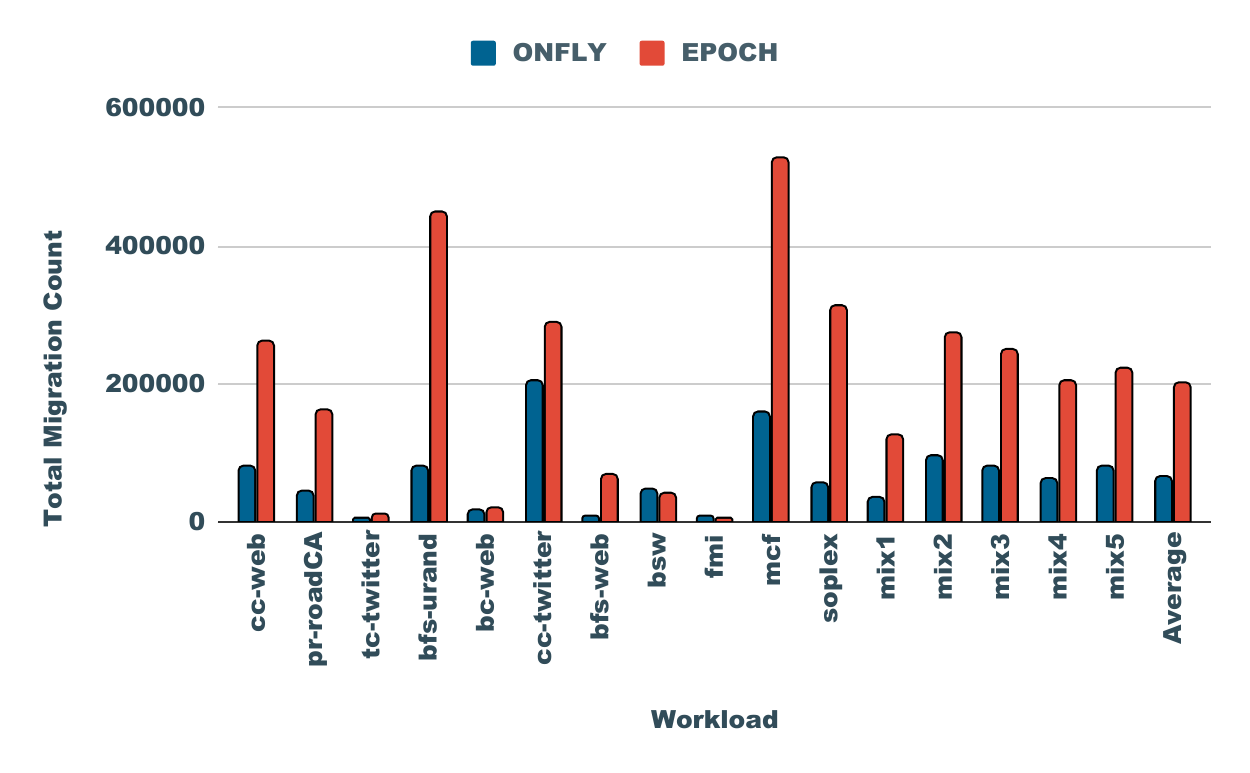}
        \caption{Total migration count of ONFLY and EPOCH.}
        \label{migr_count}
    \end{subfigure}
    \caption{(a) Normalized IPC is reported for ONFLY-DUON when compared with ONFLY, EPOCH-DUON when compared to EPOCH and ADAPT-THOLD-DUON when compared to ADAPT-THOLD, respectively. (b) Migration count is reported for a threshold value of 64 with 1GB HBM and 16GB PCM.
    }
    \label{ipc_imp_with_ept_migr_count}
    \vspace{-4mm}
\end{figure}

Fig.~\ref{ipc_imp_with_ept} shows normalized IPC improvement of ONFLY, EPOCH, and ADAPT-THOLD, when integrated with {\tt Duon}. ONFLY achieves an average IPC improvement of 1.83\% when {\tt Duon} is integrated with it. EPOCH achieves an average IPC improvement of 3.87\% indicating a consistent benefit in IPC with {\tt Duon} integration. \texttt{Duon} integration with ONFLY not only reduces overhead due to cache line invalidations and TLB shootdown, but also reduces LLC miss rate for workloads like {\it cc-web}, {\it bc-web}, {\it mcf}, {\it mix1}, {\it mix4} and {\it mix5} by 0.20\%, 0.16\%, 0.28\%, 0.17\%, 0.78\%, and 0.20\% respectively. This improvement in LLC miss rate can be attributed to the hits to the other wise invalidated (as in the non {\it Duon} cases) blocks in the cache. 

 For ONFLY technique in Fig.~\ref{migr_count}, frequent migrations do not always significantly improve IPC because they can introduce overhead without yielding proportional performance gains in some cases like \texttt{cc-twitter}, \texttt{mix1}. ONFLY migrates pages as soon as they become hot, often before their access pattern stabilize. This eagerness leads to repeated migrations of pages that may not remain hot long enough to benefit from being placed in faster memory. As a result, resources are spent on migration operation, without a corresponding reduction in access latency. Page migration overhead in ONFLY is not added in overall cycles (as explicitly mentioned in~\cite{OnTheFly}) except when there is access to the page that is under migration, because according to ONFLY proposal address reconciliation happens in the background when remap table size is 50\% full. \texttt{Duon} eliminates cache line invalidations and TLB shotdowns, but the reduced overhead doesn't show up in stats as compared to ONFLY, because in ONFLY itself cache line invalidation overhead is not added for pages if they are not accessed during address reconciliation process. On the other hand, for workloads such as \texttt{tc-twitter}, \texttt{fmi}, \texttt{soplex} having less migration count for few hot but critical pages, ONFLY-DUON shows high IPC improvement. Lastly, we observe that for workloads having high migration count under EPOCH shown in Fig.~\ref{migr_count}, {\tt Duon} shows more IPC improvement by reducing the page migration overhead.  Workloads such as \texttt{bfs-urand}, \texttt{mcf}, in Fig.~\ref{ipc_imp_with_ept_migr_count} experience high migration count for EPOCH, and hence show high IPC gain with {\tt Duon}. For ADAPT-THOLD, integration of {\tt Duon} results in an average IPC improvement of 0.91\%. While this gain is smaller compared to the other two configurations, it demonstrates that {\tt Duon} still provides measurable benefits. 


\subsection{Sensitivity Analysis}
\label{sensitivity_analysis}
This section examines the performance improvement achieved by the {\tt Duon}-integrated techniques (ONFLY-DUON and EPOCH-DUON) under varying memory configurations and threshold values. Three distinct configurations of fast and slow memories are considered:

\begin{itemize}
    \item Configuration 1: FAS with 1GB HBM and 16GB PCM
    \item Configuration 2: FAS with 256MB HBM and 16GB PCM
    \item Configuration 3: FAS with 1GB HBM and 16GB DDR4 DRAM
\end{itemize}

For every configuation, IPC improvements are compared for threshold values of 64 and 128 to evaluate the impact of memory capacity and threshold settings on performance. \\

\begin{figure}[h]
    \centering
    \begin{subfigure}{0.48\textwidth}
        \centering
        \includegraphics[width=0.8\linewidth]{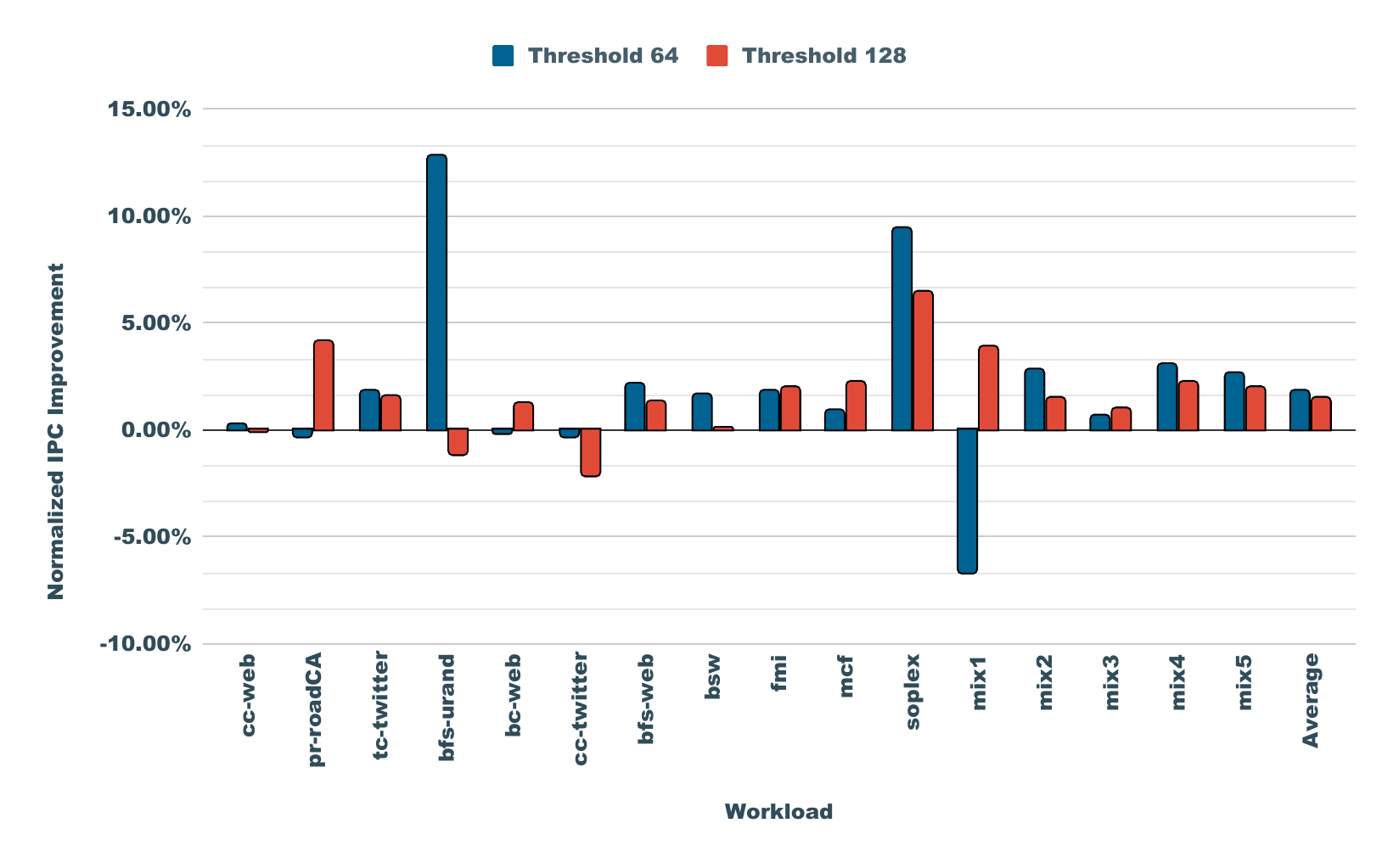}
         \caption{Normalized IPC improvement of ONFLY-DUON w.r.t. ONFLY.}
        \label{ipc_imp_hma_ept_over_onfly_th_64_128}
    \end{subfigure}
    \hfill 
    \begin{subfigure}{0.48\textwidth}
        \centering
        \includegraphics[width=0.8\linewidth]{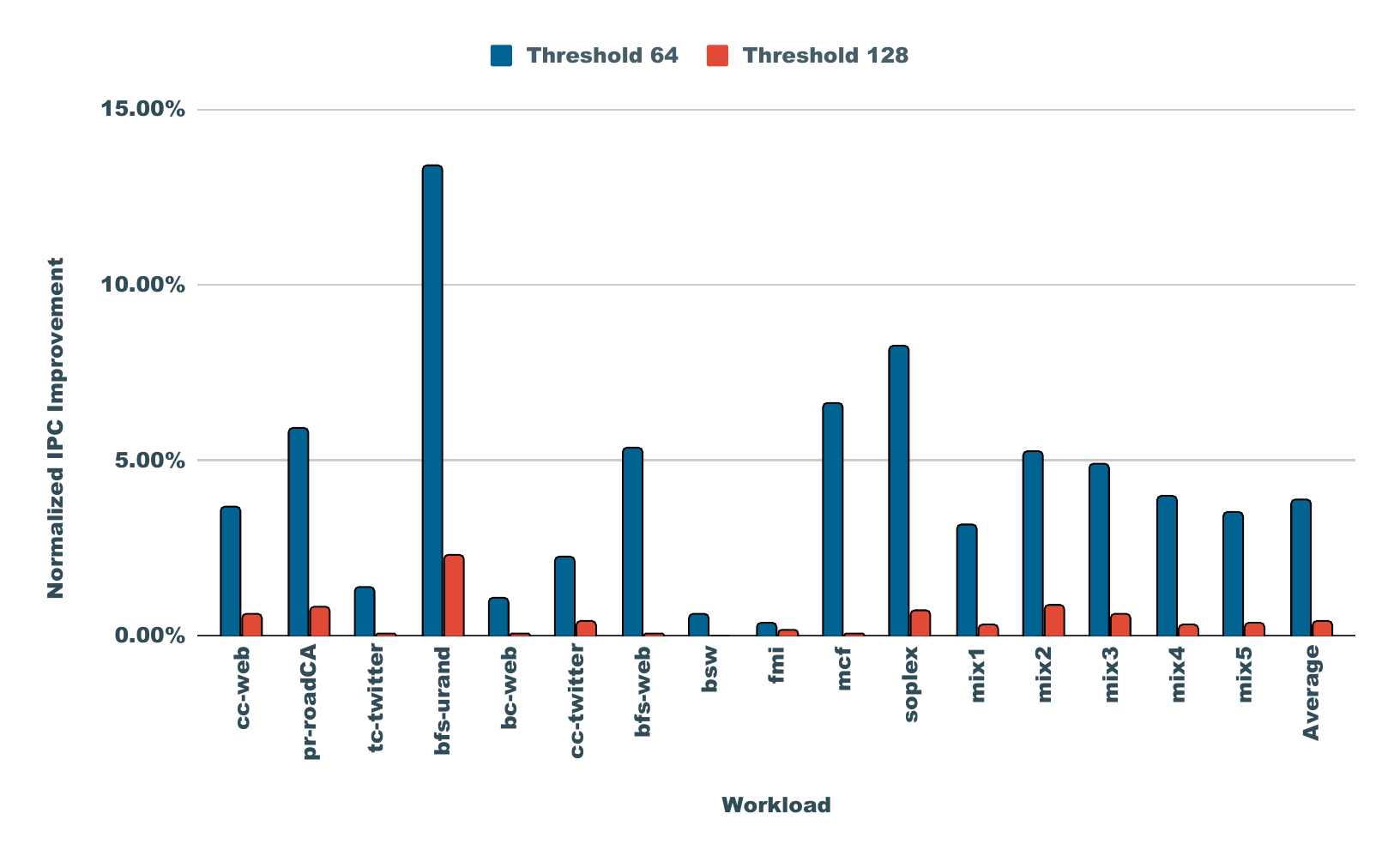}
        \caption{Normalized IPC improvement of EPOCH-DUON w.r.t. EPOCH.}
        \label{ipc_imp_hma_epoch_ept_over_epoch_th_64_128}
    \end{subfigure}
    \caption{Normalized IPC improvement with DUON for ONFLY and EPOCH, for threshold values of 64 and 128, for FAS made up of 1GB HBM and 16GB PCM.}
    \label{fig:mainfigure}
\end{figure}

\textbf{Configuration 1: FAS with 1GB HBM with 16GB PCM} \\
In this configuration, the larger HBM size of 1 GB provides greater capacity for frequently accessed data, enabling IPC improvements for the {\tt Duon}-integrated techniques as shown in Fig.~\ref{ipc_imp_hma_ept_over_onfly_th_64_128} and \ref{ipc_imp_hma_epoch_ept_over_epoch_th_64_128}. ONFLY-DUON shows normalized IPC improvement of 1.83\% for a threshold of 64, decreasing to 1.50\% for a higher threshold of 128. This trend indicates that a lower threshold value enhances the benefits of {\tt Duon} due to higher frequent migration, which can better utilize the larger HBM capacity. Similarly, EPOCH-DUON achieves its highest IPC improvement of 3.87\% for the threshold of 64, which drops significantly to 0.42\% for the threshold of 128. These results highlight the sensitivity of {\tt Duon} performance to the threshold value, particularly in scenarios with higher memory capacity. \\

\begin{figure}[h]
    \centering
    \begin{subfigure}{0.48\textwidth}
        \centering
        \includegraphics[width=0.8\linewidth]{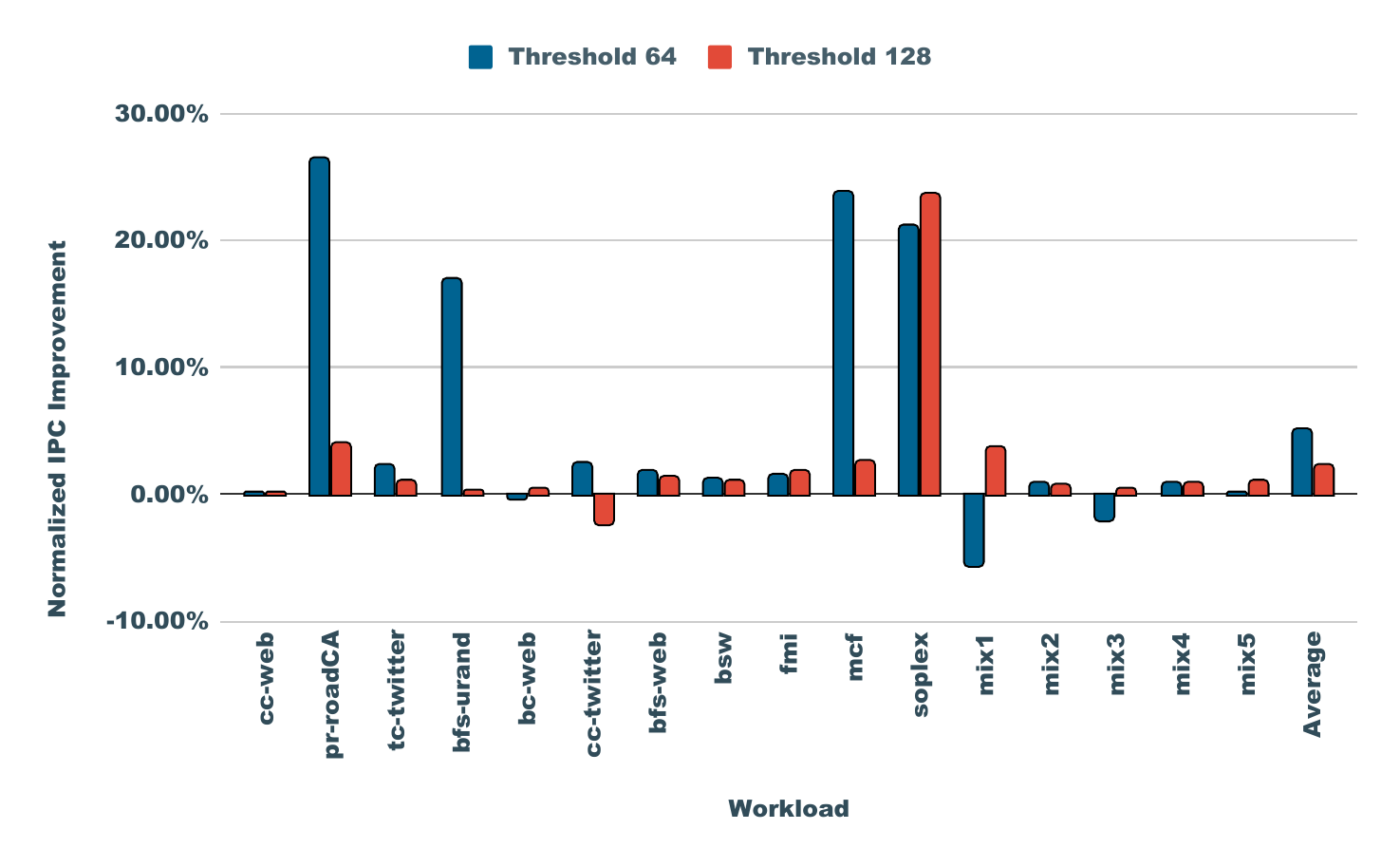}
         \caption{Normalized IPC improvement of ONFLY-DUON w.r.t ONFLY.}
        \label{ipc_imp_hma_ept_over_onfly_th_64_128_HBM256MB}
    \end{subfigure}
    \hfill 
    \begin{subfigure}{0.48\textwidth}
        \centering
        \includegraphics[width=0.8\linewidth]{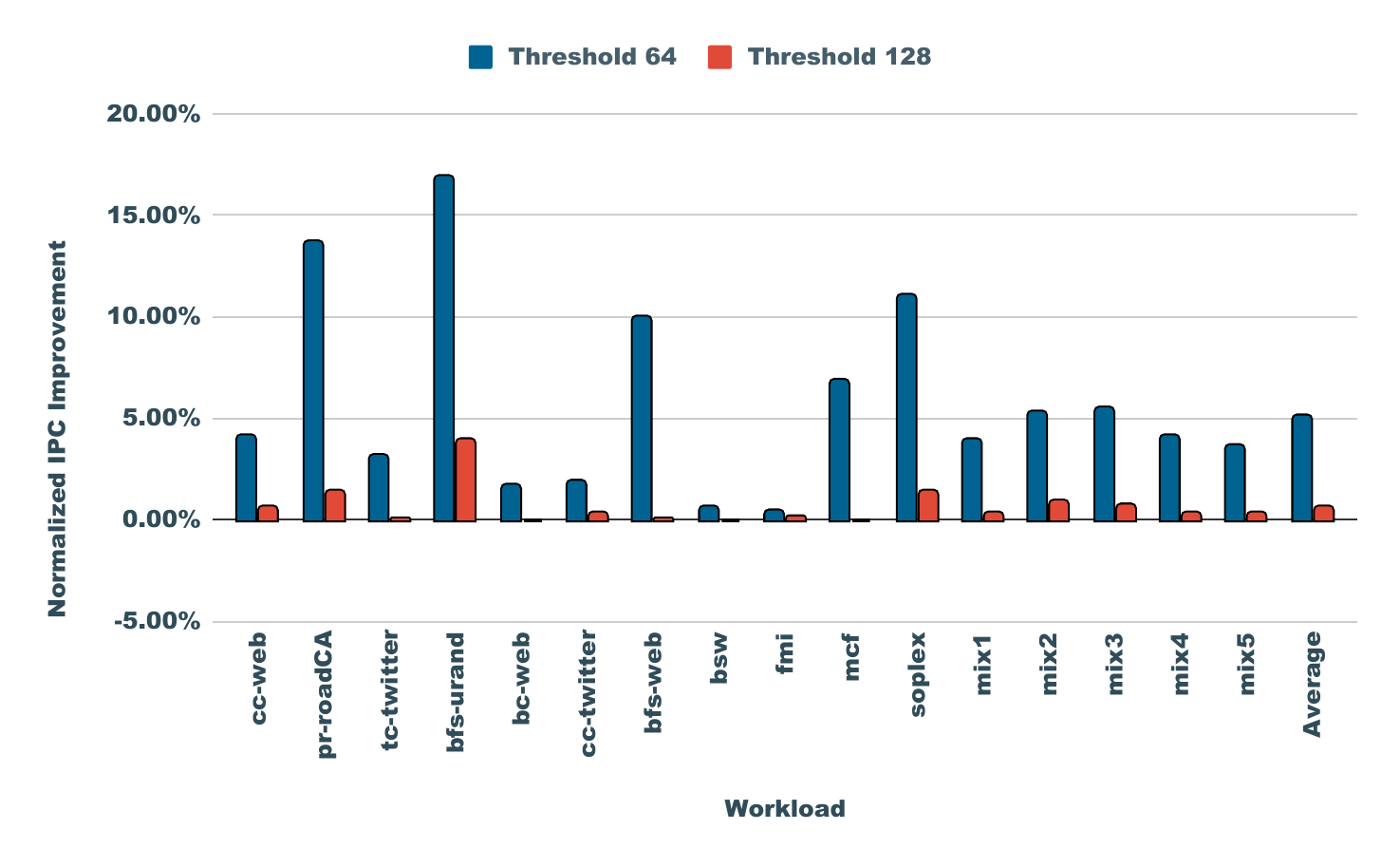}
        \caption{Normalized IPC improvement of EPOCH-DUON w.r.t. EPOCH.}
        \label{ipc_imp_hma_epoch_ept_over_epoch_th_64_128_HBM256MB}
    \end{subfigure}
    \caption{Normalized IPC improvement with DUON for ONFLY and EPOCH, for threshold values of 64 and 128, for FAS made up of 256MB HBM and 16GB PCM.}
    \label{fig:mainfigure}
\end{figure}

\textbf{Configuration 2: FAS with 256MB HBM and 16GB PCM} \\
The reduced HBM size of 256 MB imposes stricter constraints on memory management in this configuration, resulting in greater IPC improvements compared to Configuration 1 as shown in Fig.~\ref{ipc_imp_hma_ept_over_onfly_th_64_128_HBM256MB} and \ref{ipc_imp_hma_epoch_ept_over_epoch_th_64_128_HBM256MB}. ONFLY-DUON achieves a substantial IPC improvement of 5.14\% and 2.30\% for thresholds 64 and 128, respectively. The higher improvement at a lower threshold highlights the ability of {\tt Duon} to optimize performance under constrained HBM capacity by effectively avoiding the TLB shootdown and cache invalidations overheads of the migrations. For EPOCH-DUON, IPC improvement is 5.18\% at a threshold of 64, reducing to 0.61\% at a threshold of 128. These results demonstrate that DUON integration is particularly effective under lower memory capacity, with smaller thresholds allowing for better utilization of available HBM resources. 

\begin{figure}[h]
\vspace{-3mm}
  \centering
  \includegraphics[width=0.4\linewidth, keepaspectratio, height=5cm]{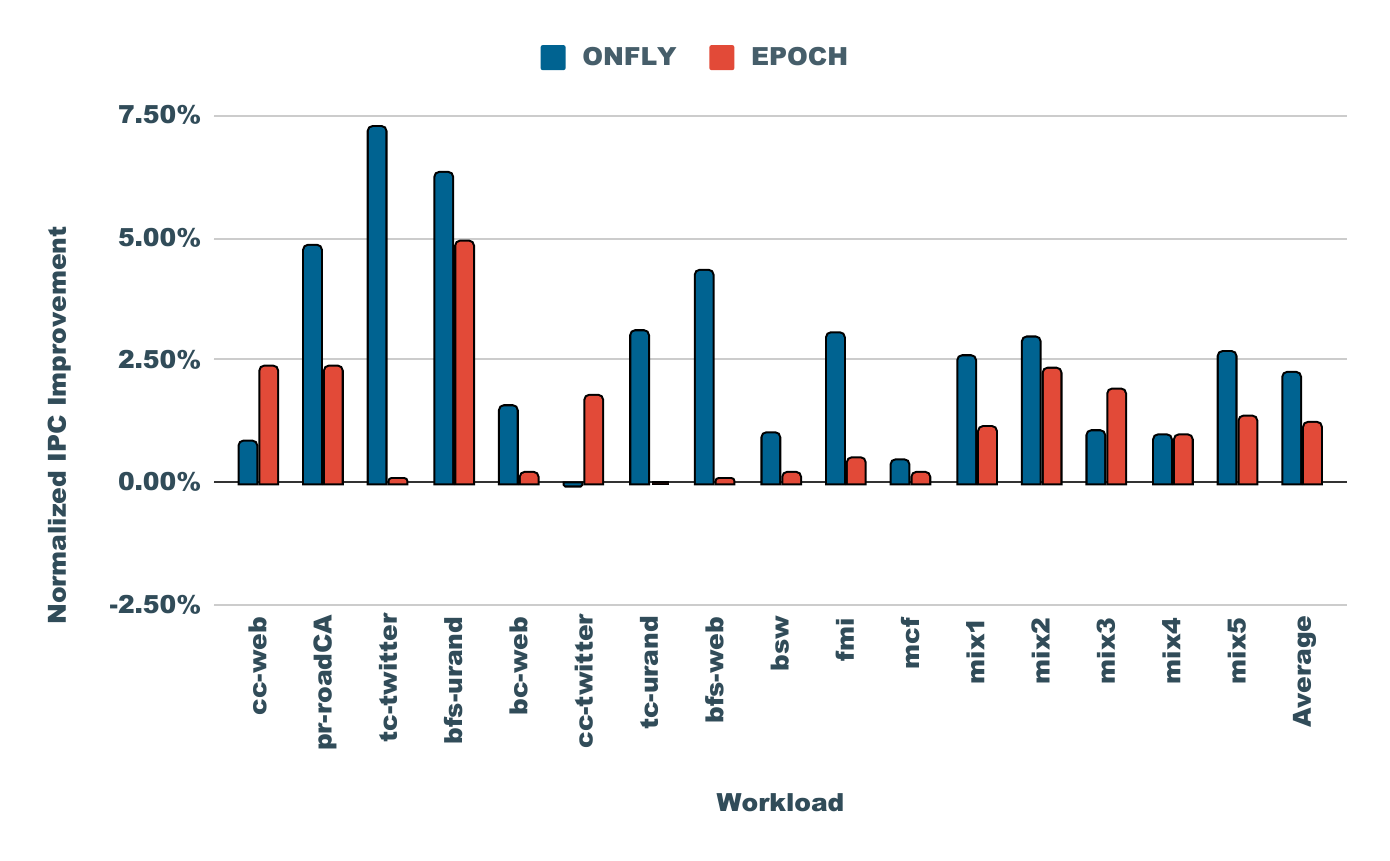}
  \caption{Normalized IPC improvement of ONFLY and EPOCH using DUON, with threshold value of 128,for FAS made up of 1GB HBM and 16GB DDR4 DRAM.}
  \label{hbm_ddr_ipc_imp_with_ept}
  \Description{Fig. shows Normalized IPC Improvement of ONFLY and EPOCH using DUON}
  \vspace{-3mm}
\end{figure}

\textbf{Configuration 3: FAS with 1GB HBM and 16GB DDR4 DRAM} \\
In this configuration, we study the impact of substituting the slower PCM memory with a faster DDR4 DRAM, to evaluate the architectural flexibility and performance impact. The results in Fig.~\ref{hbm_ddr_ipc_imp_with_ept} demonstrate that DUON provides a normalized IPC improvement of 2.25\% and 1.24\% when used with the existing ONFLY and EPOCH techniques, highlighting its efficiency and adaptability across varying memory configurations. \\

The sensitivity analysis reveals a significant dependency of IPC improvement on both memory capacity and threshold values. Across all the three configurations, lower threshold (64) consistently yield higher IPC improvements, indicating that frequent page migrations facilitated by smaller thresholds are effectively supported by \texttt{Duon}. Furthermore, the larger IPC improvements observed in configuration 2 with 256MB HBM highlight the effectiveness of DUON in scenarios with limited fast memory, where efficient migration and memory access are critical. Overall, these findings emphasize the importance of tuning threshold values to balance migration frequency and memory utilization, particularly in systems with varying memory configurations. The integration of DUON with existing techniques proves to be a versatile solution, capable of adapting to diverse architectural constraints and performance requirements.

\subsection{Hardware cost analysis}
{\tt Duon} requires an additional storage overhead of 13.69 MB to accommodate the extended page table (shown in Fig.~\ref{ext_pt_struct}), which incorporates remapped physical address and migration metadata to support efficient page migration. This extended page table is implemented across a memory hierarchy consisting of 1 GB of faster memory and 16 GB of slower memory, with 4KB page size, accounting for total number of pages as 262144 and 4194304, respectively. Each faster and slower memory page has an additional overhead of 22 and 26 bits for metadata described in Fig.~\ref{EPT_structure}. Precisely, the remapped physical address require 18 and 22 bits as per faster and slower memory capacity respectively, migrated flag, ongoing migration flag, pair and buffer residency require 1 bit each. Given that {\tt Duon} requires an extension of TLB (shown in Fig.~\ref{ext_tlb_struct}) to store remapped physical address and page migration metadata, an additional 12.5 KB is required considering 4096 TLB entries. This leads to 29\% additional  storage with extended TLB w.r.t conventional TLB that intially requires 30.5 KB storage. Further, the extended page table overhead accounts for 0.08\% of the total main memory capacity, reflecting a minimal impact on overall memory usage while enabling advanced functionality.  In addition to the page table storage, the system includes additional ONFLY components such as Migration Controller, Hot and Cold Buffers, Wait Queue, Migration Queue, as listed in \cite{OnTheFly} except the remap table, as it is replaced by our Extended Page Table and TLB, that further support efficient page migration and memory management.


\section{Conclusion}
\label{conclude}
This research proposes a novel approach to memory page relocation through {\tt Duon}. By preserving initial virtual-to-physical mappings and using a suite of metadata flags, the proposed methodology mitigates the performance overhead associated with traditional page migration. The incorporation of extended TLB and page table along with features such as the ongoing migration flag, hot and cold buffers, bit vector further optimizes the migration process, ensuring efficient handling of cache lines and memory requests. Our evaluation demonstrates significant performance gains, with a maximum IPC improvement of 13.39\% and an average improvement of 3.87\% across diverse workloads. These results highlight the system's ability to consistently enhance efficiency and optimize performance in varied scenarios. As computing systems grow increasingly complex, {\tt Duon} provides a robust and scalable solution to memory management due to page migration challenges, paving the way for enhanced performance and reliability in future architectures.




\bibliographystyle{ACM-Reference-Format}
\bibliography{sample-base}

\end{document}